\documentclass[lettersize, journal]{IEEEtran}
\usepackage[utf8]{inputenc} 
\usepackage[T1]{fontenc}
\usepackage{url}
\usepackage{ifthen}
\usepackage[cmex10]{amsmath}
\usepackage{array} 

\usepackage{enumerate}
\usepackage{amssymb}
\usepackage{amsmath} 

\usepackage{algorithm}
\usepackage{algorithmicx}
\usepackage{algpseudocode}

\usepackage{mathrsfs}
\usepackage{bm}
\usepackage{tikz}
\usetikzlibrary{arrows}
\usepackage{subfig}
\usepackage{graphicx,graphics, booktabs,multirow}
\usepackage{epstopdf}
\usepackage{multirow}
\usepackage{tabularx} 
\usepackage{booktabs}

\DeclareMathOperator*{\argmin}{arg\,min}
\usepackage{pifont}

\definecolor{colorhkust}{RGB}{20,43,140}
\definecolor{colortsinghua}{RGB}{116,52,129}
\definecolor{color1}{RGB}{128,0,0}

\usepackage{caption}
\usepackage{amsthm}
\usepackage{cite}
\newtheorem{thm}{Theorem}
\newtheorem{lem}{Lemma}

\newtheoremstyle{mynote}%
  {}%
  {}%
  {}%
  {}%
  {\bfseries}%
  {}%
  {.5em}%
  {}%
\theoremstyle{mynote}
\newtheorem{defn}{Definition}

\theoremstyle{remark}

\usepackage{makecell}

\begin{document}

        \title{Low-Complexity CSI Feedback for FDD Massive MIMO Systems via Learning to Optimize}

      \author{Yifan Ma, \textit{Graduate Student Member}, \textit{IEEE},  Hengtao He, \textit{Member}, \textit{IEEE}, Shenghui Song, \textit{Senior Member}, \textit{IEEE}, Jun Zhang \textit{Fellow}, \textit{IEEE}, and Khaled B. Letaief, \textit{Fellow}, \textit{IEEE}
      	\thanks{     		
      	The authors are with the Department of Electronic and Computer Engineering, The Hong Kong University of Science and Technology, Hong Kong (E-mail: \{ymabj, eehthe, eeshsong, eejzhang, eekhaled\}@ust.hk).  
}
}
        
        \maketitle
        
\begin{abstract}
In frequency-division duplex (FDD) massive multiple-input multiple-output (MIMO) systems, the growing number of base station antennas leads to prohibitive feedback overhead for downlink channel state information (CSI). To address this challenge, state-of-the-art (SOTA) fully data-driven deep learning (DL)-based CSI feedback schemes have been proposed. However, the high computational complexity and memory requirements of these methods hinder their practical deployment on resource-constrained devices like mobile phones.
To solve the problem, we propose a model-driven DL-based CSI feedback approach by integrating the wisdom of compressive sensing and learning to optimize (L2O). Specifically, only a linear learnable projection is adopted at the encoder side to compress the CSI matrix, thereby significantly cutting down the user-side complexity and memory expenditure. 
On the other hand, the decoder incorporates two specially designed components, i.e., a learnable sparse transformation and an element-wise L2O reconstruction module. The former is developed to learn a sparse basis for CSI within the angular domain, which explores channel sparsity effectively. The latter shares the same long short term memory (LSTM) network across all elements of the optimization variable, eliminating the retraining cost when problem scale changes. 
Simulation results show that the proposed method achieves a comparable performance with the SOTA CSI feedback scheme but with much-reduced complexity, and enables multiple-rate feedback.

\end{abstract}

\section{Introduction}
Massive multiple-input multiple-output (MIMO) is regarded as a key enabler for the fifth-generation and beyond wireless communication systems, as it empowers high throughput, simultaneous multiple streams, and ubiquitous coverage for diverse applications \cite{FiveG}. For future sixth-generation (6G) wireless communication networks, extremely large-scale MIMO is considered as a critical technological advancement, where a much larger number of antennas will be deployed at the base station (BS) \cite{wang2023tutorial, wang6G}. However, such large-scale MIMO systems pose significant challenges in the physical layer algorithm design. For example, 
in frequency-division duplexing (FDD) massive MIMO systems, accurate downlink channel state information (CSI) needs to be fed from users back to the BS for high-quality downlink beamforming. 
Unfortunately, the dimension of the CSI escalates substantially with the number of antennas at the BS, resulting in a prohibitive feedback overhead if the full CSI matrix is directly sent back. The conventional compressive sensing (CS)-based methods, widely applied for CSI compression and recovery \cite{Sparsity, Vince}, suffer from noteworthy limitations, such as the impractical assumption of channel sparsity, the limited ability to exploit the channel structures, and the high computational cost of the iterative operations \cite{ma2023lightweight}. Therefore, innovative technologies are imperative to solve the high-dimensional nonlinear CSI feedback problem.

With the success of artificial intelligence (AI) in various fields, its integration with wireless communication has attracted significant interests recently \cite{Khaled19Roadmap}. One key application of deep learning (DL) in the physical layer is DL-based CSI feedback \cite{OverviewGuo}, which leverages auto-encoder and decoder structures to compress and reconstruct the downlink CSI. These kinds of fully data-driven CSI feedback methods outperform traditional algorithms in terms of performance \cite{CsiNet, ConvCsiNet, TransNet}, thus attracting widespread attention from both the academic and industry. 
The authors of \cite{CsiNet} proposed a convolutional neural network (CNN)-based scheme, named CsiNet, which outperforms the CS-based algorithms especially with low compression ratios. Several subsequent studies, including ConvCsiNet \cite{ConvCsiNet} and TransNet \cite{TransNet}, aimed to further improve the feedback accuracy using deeper CNNs and attention mechanism, respectively. However, the performance improvement is achieved at the cost of computational complexity. For example, the number of floating point operations (FLOPs) of TransNet is almost 11 times more than that of CsiNet. 
Although TransNet achieves the state-of-the-art (SOTA) performance for CSI feedback, its heavy computational complexity and memory cost in the encoder side hinder the practical deployment on resource-constrained devices, such as mobile phones, internet-of-things (IoT) devices, and embedded systems \cite{tang2018FLOPs}. 

Existing auto-encoder and decoder-based CSI feedback schemes are completely data-driven, and thus ignore the physical characteristics of the wireless channel in the encoding and decoding process. This typically leads to a large number of learnable parameters, tricky training schemes, and can also drag down their performance without explicit physical guidance \cite{Zhang23Unroll, Ma22NC}. To solve these issues, another line of research combines communication domain knowledge with DL, where deep unfolding is considered as one of the representative solutions \cite{Monga21Unroll, He19Unfold, He20Unfold, Wentao23DEQ}. 
Deep unfolding relates iterative optimization methods with deep neural networks. It treats each iteration in the optimization algorithm as one layer of the neural network, where a number of trainable parameters are introduced to be learned by DL techniques. In deep unfolding-based CSI feedback approaches, the CS processing pipeline is preserved, i.e., a small number of codewords (observations) are obtained through linear projection and nonlinear learnable mappings are adopted to recover the CSI. For example, the authors of \cite{Wang20LISTA} proposed a sparse autoencoder to learn the sparse transformations in each iteration of iterative shrinkage-thresholding algorithm (ISTA). In \cite{Hu24LORA}, instead of using $l_{1}$-norm as the regularization term, a learnable regularization module is introduced in ISTA to automatically adapt to the characteristics of CSI. Those proposals adopt a single linear projection at the encoder side, making it applicable for resource-constrained devices in practice. However, traditional deep unfolding methods are built by truncating an iterative algorithm into finite and fixed layers, which makes it difficult to scale to variable numbers of iterations and hard to ensure convergence \cite{Wentao23DEQ}.
Additionally, the direct parameterization requires dimension matching of learnable parameters and the problem scale, indicating that the model, once trained, is not applicable to optimization problems of varying scales during inference \cite{Yin23MS4L2O}. 
For massive MIMO CSI feedback, the compression ratio has to be adjusted according to the dynamic environments and varying coherence time \cite{Guo20MultiRate}. Therefore, it is crucial to develop a DL-based CSI feedback scheme that guarantees convergence and is able to generalize to different compression ratios.

To address these challenges, in this paper, we propose a model-driven DL method for CSI feedback. Inspired by the recent success of utilizing AI, especially DL, for solving mathematical problems, we propose a Learning to Optimize (L2O)-based approach that combines the wisdom of both the CS algorithm and DL. Using L2O models to solve optimization problems involves the design of a learnable update rule \cite{chen2022learning, andrychowicz2016learning, Yin23MS4L2O}, leading to an autonomously learned optimization algorithms from data. 
While L2O strategies can achieve a faster convergence and better performance than conventional non-learning optimization algorithms \cite{andrychowicz2016learning, Yin23MS4L2O}, directly implementing them for CSI feedback still meets obstacles. Specifically, the reconstruction performance is highly dependent on the signal sparsity of the data in a specific transform domain. However, the wireless channel is not exactly sparse in some domains. Without an effective transformation and sufficient sparsity level, the L2O method will have the severe performance degradation. Although traditional manually designed transformations, e.g., discrete fourier transform (DFT) and wavelet transformation, can be utilized,  they require a large number of iterations at the decoder, resulting in high computational complexity and restricting their practical applications. Therefore, it requires special
design for the L2O-based CSI feedback approach.

\subsection{Contributions}
To deal with the imbalanced computational capability between the mobile equipment and BS and reduce the retraining cost when problem scale changes, we propose an L2O-based CSI feedback scheme, i.e., Csi-L2O, in this paper. It enjoys ultra low-complexity at the encoder side, comparable performance compared to SOTA, and adaptability to multiple feedback rates without retraining the neural network. 
The major contributions are summarized as follows:

\begin{itemize}
	\item \emph{Low Complexity:} The overall framework integrates the wisdom of CS and DL. Inspired by CS, the codeword is obtained through a linear projection at the user side and full CSI is recovered via a parameterized update rule at the BS side. Different from the auto-encoder and decoder structures that adopt convolutional kernels, fully connected layers, or attention mechanism, the linear projection encoding module inherently enjoys ultra low-complexity, which is more suitable for practical wireless communication systems.
	
	\item \emph{Comparable Performance:} To maintain performance, we propose a data-driven channel sparse transformation and L2O module at the decoder side.  
In contrast to manually designed sparse transformation, we propose to learn the sparse transformation in the angular domain, resulting in a more efficient sparse representation for CSI. The following L2O module is proposed to capture dynamics among different layers and learns the optimization update rule automatically from data, ensuring a good reconstruction accuracy. 

	\item \emph{No Retraining Cost:} To make the proposed Csi-L2O generalizable to different compression ratios, we adopt an ``element-wise” long short term memory (LSTM) to generate the optimization parameters at the decoder. In particular, the same neural network is shared across each element of the optimization variables, so that the proposed single model can be applied to optimization problems of any scales without retraining and enable the multiple-rate feedback.

	\item \emph{Simulations:} Extensive simulations will demonstrate that the performance of the proposed L2O-based method is close to existing SOTA, i.e., TransNet, while enjoying significantly improved computational efficiency compared with the fully data-driven methods. In particular, the proposed L2O method achieves 3.88 dB higher reconstruction accuracy than SOTA, TransNet, in an indoor scenario with a compression ratio of $1/16$. In addition, the encoder side FLOPs of the proposed method is only $0.15\%$ of that of SOTA, making the deployment to resource constraint devices practical. 
\end{itemize}

\subsection{Organization and Notations} 
The paper is organized as follows. Section \uppercase\expandafter{\romannumeral2} introduces the system model and existing approaches. In Section \uppercase\expandafter{\romannumeral3}, the key design properties and the proposed Csi-L2O architecture are presented. Then, we perform the convergence analysis and computational complexity analysis in Section \uppercase\expandafter{\romannumeral4}. Extensive simulations are demonstrated in Section \uppercase\expandafter{\romannumeral5} and conclusions are drawn in Section \uppercase\expandafter{\romannumeral6}.

In this paper, $x$ is a scalar, ${\mathbf{x}}$ is a vector, and ${\mathbf{X}}$ denotes a matrix. 
Let ${{\mathbf{X}}^T}$ and ${{\mathbf{X}}^H}$ denote the transpose and conjugate transpose of matrix ${\mathbf{X}}$, respectively.  
${{\mathbf{I}}}$ stands for an identity matrix, $\mathbf{1}$ represents the vector whose all elements are all ones, and $\mathbf{0}$ denotes the zero vector.
$|| {\mathbf{X}} ||_2$ and $\mathbf{X}^{-1}$ denote the Frobenius norm and the inverse of matrix $\mathbf{X}$, respectively.
$\mathbb{E}\{{\cdot}\}$ denotes the statistical expectation.
$f_\mathbf{\theta}$ denotes a mapping parameterized by learnable parameters $\mathbf{\theta}$.
Function $\operatorname{sign}(\cdot)$ represents element-wise sign fuction. Function $\max(\mathbf{x}, \mathbf{y})$ returns element-wise maximum value between vector $\mathbf{x}$ and $\mathbf{y}$.
$\mathbf{X} = \text{diag}(\mathbf{x})$ defines $\mathbf{X}$ as a diagonal matrix with $\mathbf{x}$ as its diagonal. 
$\mathbb{C}^{m \times n}$ is the set of all ${m \times n}$ complex-valued matrices. 
The Hadamard product is denoted by $\odot$.

\section{System Model and Existing Approaches} \label{sec:sys_model}
In this section, we first formulate the CSI feedback problem. Then, existing DL-based CSI feeback schemes are introduced, which motivates the proposed method.
\subsection{System Model}

\begin{figure}[t] 
\centering
\includegraphics[width=0.5\textwidth]{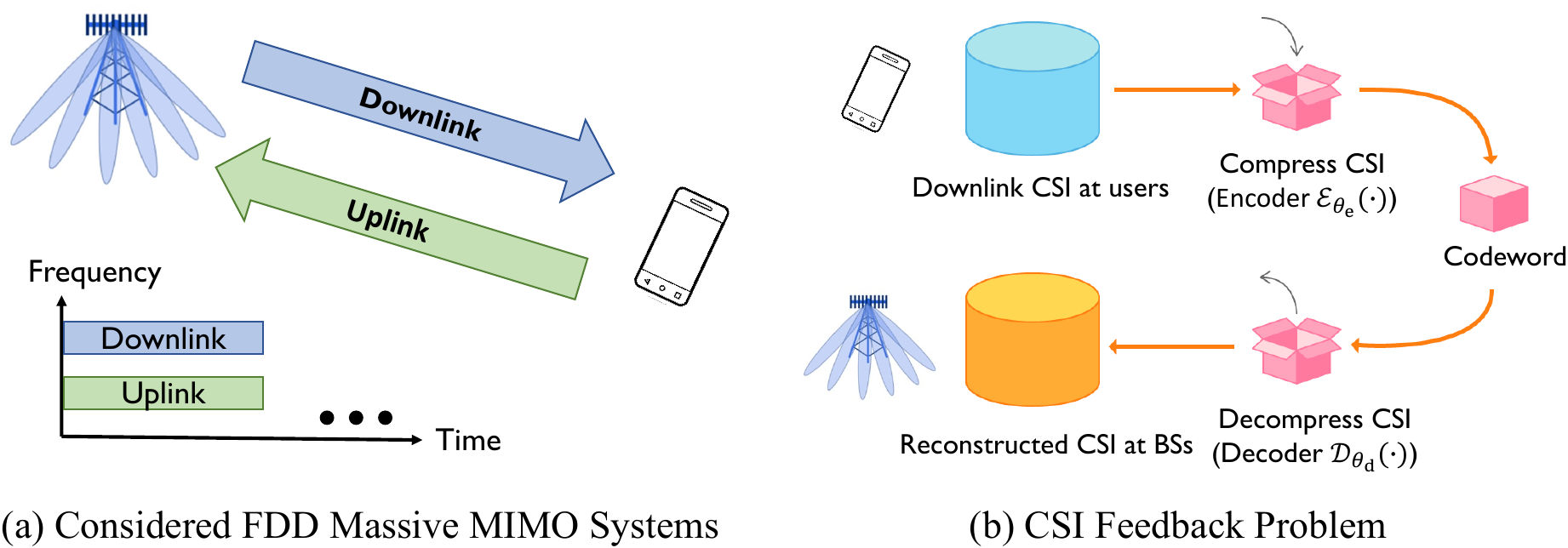} 
\caption{An illustration of the considered communication system and CSI feedback problem.} 
\label{SystemModels} 
\end{figure}

As illustrated in Fig. \ref{SystemModels}(a), we consider a single-cell FDD massive MIMO system where the BS is equipped with $N_t$ antennas and the user is equipped with a single antenna. For ease of illustration, a single user case is considered while the proposed scheme can be easily generalized to the multi-user scenario. An orthogonal frequency division multiplexing (OFDM) system with $N_c$ subcarriers is considered. The received signal on the $n$-th subcarrier is expressed as
\begin{equation}
y_n = \mathbf{h}_n^H \mathbf{v}_n x_n + z_n,
\end{equation}
where $\mathbf{h}_n \in \mathbb{C}^{N_t \times 1}$, $\mathbf{v}_n \in \mathbb{C}^{N_t \times 1}$, $x_n \in \mathbb{C}$, and $z_n \in \mathbb{C}$ denote the downlink channel vector, the downlink beamforming vector, the transmit symbol, and the additive noise of the $n$-th subcarrier, respectively. The downlink beamforming requires the BS to know the downlink CSI, denoted by $\mathbf{H} = [\mathbf{h}_1, \cdots, \mathbf{h}_{N_c}]^H \in \mathbb{C}^{N_c \times N_t}$. In this paper, we assume that the downlink channel is perfectly known at the user side via pilot-based training and focus on the efficient feedback design \cite{CsiNet, ConvCsiNet, TransNet}.

The channel matrix $\mathbf{H}$ contains $2 N_c N_t$ real elements. As $N_c$ and $N_t$ are large in FDD massive MIMO systems, directly feeding back $\mathbf{H}$ will result in prohibitive feedback overhead. To tackle this issue, we first sparsify $\mathbf{H}$ in the angular-delay domain using a 2D discrete Fourier transform (2D-DFT) \cite{CsiNet} as follows
\begin{equation}
\mathbf{H}' = \mathbf{F}_\mathrm{d} \mathbf{H} \mathbf{F}_\mathrm{a},
\end{equation}
where $\mathbf{F}_\mathrm{d} \in \mathbb{C}^{N_c \times N_c}$ and $\mathbf{F}_\mathrm{a} \in \mathbb{C}^{N_t \times N_t}$ are two DFT matrices. Only the first $N_a$ rows of $\mathbf{H}'$ contain significant values and other elements are close to zero because the time delays between multipath arrivals are within a limited period \cite{CsiNet}. Therefore, we take the first $N_a$ rows of $\mathbf{H}'$ ($N_a < N_c$) and define a new matrix $\mathbf{H}'' \in \mathbb{C}^{N_a \times N_t}$. By doing this, we can compress $\mathbf{H}''$ instead of $\mathbf{H}$ with only $2N_a N_t$ elements and imperceptible information loss. 

DL-based methods have been applied for CSI feedback \cite{CsiNet, ConvCsiNet, TransNet}. As demonstrated in Fig. \ref{SystemModels}(b), the encoding process at the user side is given by
\begin{equation} \label{encoder}
\mathbf{s} = \mathcal{E}_{\theta_\mathrm{e}}(\mathbf{H}''),
\end{equation}
which further compresses the channel matrix $\mathbf{H}''$ into an $M \times 1$ codeword $\mathbf{s}$. The parameterized mapping $\mathcal{E}_{\theta_\mathrm{e}}(\cdot)$ denotes the compression procedure and $\theta_\mathrm{e}$ is the trainable parameters in the encoder. The compression ratio is defined as $M/2N_a N_t$. 
We use the same setting as \cite{CsiNet, ConvCsiNet, TransNet} and assume $\mathbf{s}$ is sent to the BS via error-free transmission.
After receiving the codeword, the BS reconstructs the channel matrix through a decoder, expressed as
\begin{equation} \label{decoder}
\hat{\mathbf{H}}'' = \mathcal{D}_{\theta_\mathrm{d}}(\mathbf{s}),
\end{equation}
where $\mathcal{D}_{\theta_\mathrm{d}}(\cdot)$ denotes the recovery procedure and $\theta_\mathrm{d}$ represents the trainable parameters at the decoder. The objective of the CSI feedback is to minimize the mean-squared-error (MSE) between the recovered channel and the true channel, given by
\begin{equation} \label{general}
\min_{\theta_\mathrm{e}, \theta_\mathrm{d}} \quad \mathbb{E} \left\{||\mathbf{H}'' - \mathcal{D}_{\theta_\mathrm{d}}(\mathcal{E}_{\theta_\mathrm{e}}(\mathbf{H}''))||_2^2 \right\}.
\end{equation}
Many existing works aim to solve Problem \eqref{general} and the most representative approaches are fully data-driven methods and deep unfolding.

\subsection{Existing Approaches}
\subsubsection{Fully Data-Driven DL-based Methods} In order to solve Problem \eqref{general}, fully data-driven DL-based methods \cite{CsiNet, ConvCsiNet, TransNet} have been developed. The mapping $\mathcal{E}_{\theta_\mathrm{e}}(\cdot)$ and $\mathcal{D}_{\theta_\mathrm{d}}(\cdot)$ can be instantiated as DL-based encoder and decoder, and jointly trained via end-to-end learning \cite{CsiNet, ConvCsiNet, TransNet}. Fully data-driven DL-based approaches obtain better performance than traditional CS-based methods, especially at low compression ratios. This is because of the powerful representation ability and universal approximation of neural networks. However, most of the existing works improve the reconstruction accuracy at the cost of higher neural network complexity, e.g., larger kernels, deeper neural networks, or complicated attention mechanism, which is not affordable for resource-constrained devices, e.g., mobile phones. For example, assume that the compression ratio is 1/16 and the CSI feedback and recovery period is 1 millisecond. The computational overhead required by the TransNet encoder is about 17.07 G floating point operations per second (FLOPS). Note that Kirin 659, one of the mid-end mobile systems on chip (SoC), has a total peak computation capability of 57.6 G FLOPS \cite{SoC}. 
If the TransNet is deployed in practice, around $30\%$ of the mobile’s computational power is used for CSI feedback, which cannot be acceptable. Although TransNet achieves SOTA performance, the extensive computational demands and memory requirements hinder its practical deployments.

\begin{figure*}[tbh]
\begin{centering}
\subfloat[\label{general:Arc}][The overall architecture of the proposed Csi-L2O.]{\begin{centering}
\includegraphics[width=0.55\textwidth]{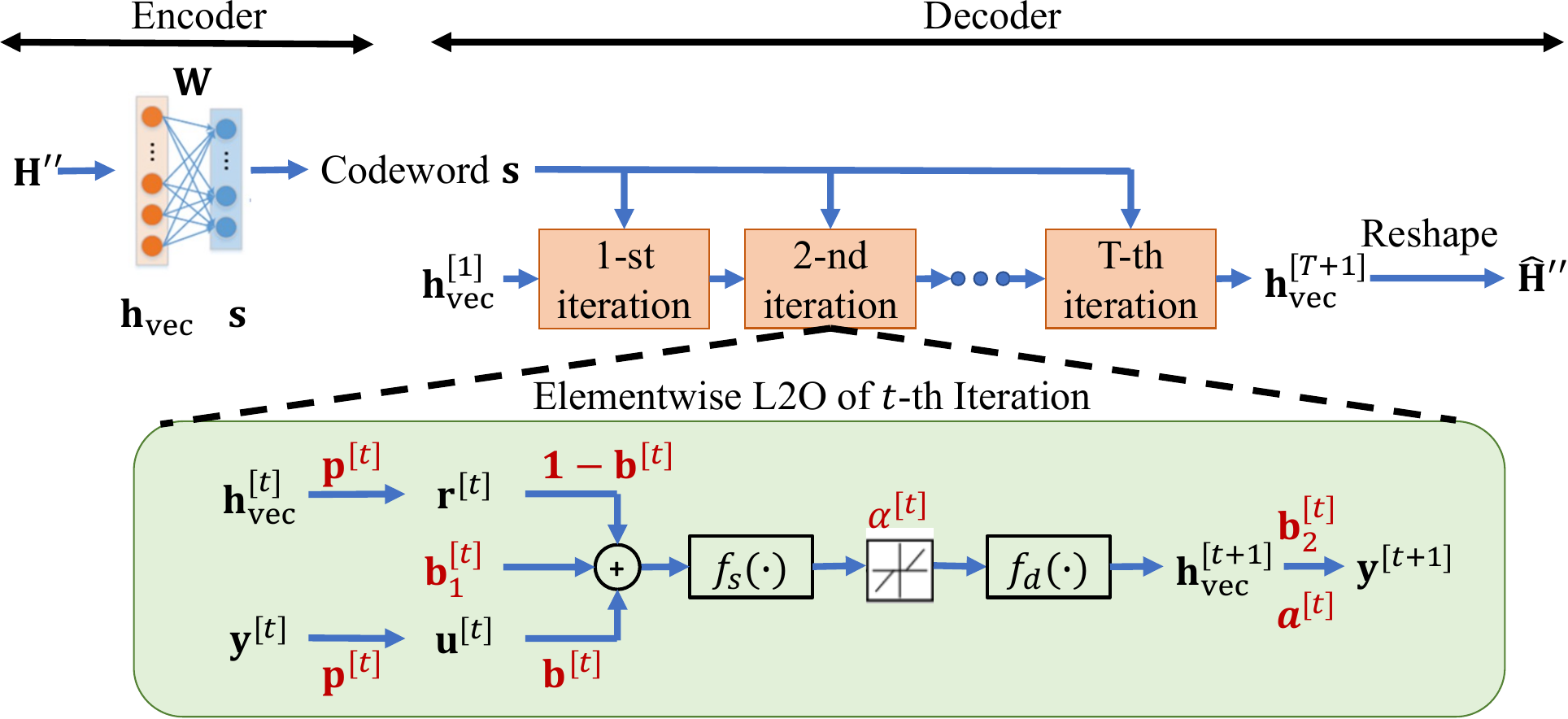}
\par\end{centering}
}\subfloat[\label{Elementwise-L2O}][The structure of the proposed element-wise L2O mechanism.]{\begin{centering}
\includegraphics[width=0.35\textwidth]{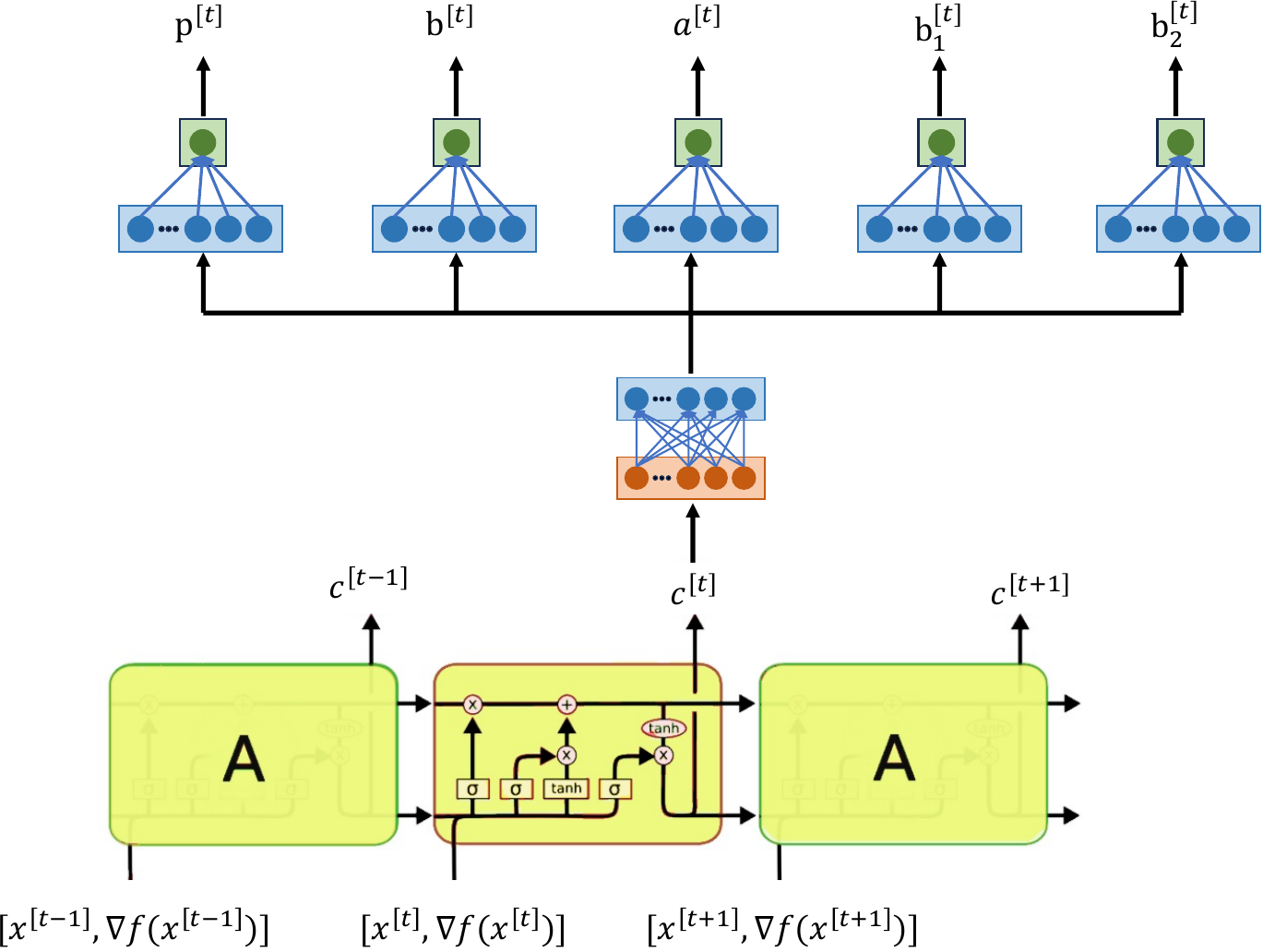}
\par\end{centering}
}\caption{(a) The overall architecture of the proposed Csi-L2O. (b) The structure of the proposed element-wise L2O mechanism. \label{Architecture}}
\par\end{centering}
\end{figure*}

\subsubsection{Deep Unfolding} By taking the physical meaning of encoding and decoding process into consideration, deep unfolding methods were proposed for CSI feedback \cite{Wang20LISTA, Hu24LORA, Xue22FISTA}. It is shown in classic CS theory that when a signal exhibits a certain sparsity in a specific transform domain, we can obtain a small number of codewords (observations) through linear projection and use nonlinear recovery mapping to get an accurate estimation of the original signal \cite{Zhang23Unroll}. By amalgamating the CS knowledge, deep unfolding-based methods implement a linear learnable encoding process to reduce the signal dimension at the encoder side. The projected codeword can be expressed as
\begin{equation} \label{compress}
\mathbf{s} = \mathbf{W} \mathbf{h_\text{vec}},
\end{equation}
where $\mathbf{W}$ is the sampling matrix and $\mathbf{h_\text{vec}} \in \mathbb{R}^{2N_a N_t \times 1}$ is the vectorized channel matrix $\mathbf{H}''$ stacking the real and imaginary part.
The decoding process at the BS can be regarded as solving an inverse problem. The dimensionality reduction in \eqref{compress} makes the signal recovery notably ill-posed. A regularization term is typically added to the optimization function to make use of known prior information about the optimal solution, which is expressed as
\begin{equation} \label{regularizer}
\min_{\mathbf{x}} \frac{1}{2} ||\mathbf{s} - \mathbf{Wx}||_2^2 + R(\mathbf{x}),
\end{equation}
where $R(\mathbf{x})$ is the regularization term. 
Typically, $l_{1}$-norm is utilized as a regularizer, i.e., $R(\mathbf{x}) = \lambda ||\Psi \mathbf{x}||_1$, where $\Psi$ is a certain sparse transformation. Problem \eqref{regularizer} is then written as 
\begin{equation} \label{Lasso}
\min _{\mathbf{x}} \frac{1}{2}\|\mathbf{s}-\mathbf{W} \mathbf{x}\|_2^2+\lambda\|\Psi\mathbf{x}\|_1.
\end{equation}

Iterative Shrinkage-Thresholding Algorithm (ISTA) is a classic iterative method to solve Problem \eqref{Lasso}, and the follow-up model-driven DL methods for CSI feedback are inspired by ISTA-based algorithms. At the $t$-th step of ISTA, the iterative process is expressed as
\begin{equation} \label{ISTA}
\begin{aligned}
\mathbf{u}^{[t]} & =\mathbf{x}^{[t-1]}-\alpha \mathbf{W}^T\left(\mathbf{W} \mathbf{x}^{[t-1]}-\mathbf{s}\right), \\
\mathbf{x}^{[t]} & =\operatorname{sign}\left(\Psi \mathbf{u}^{[t]}\right) \max \left(\mathbf{0},\left|\Psi \mathbf{u}^{[t]}\right|-\mathbf{\theta}\right),
\end{aligned}
\end{equation}
where $\mathbf{u}^{[t]}$, $\alpha$, and $\theta$ are the intermediate variable, step size, and thresholding parameter, respectively. 
In \cite{zhang2018ista}, a model-driven DL method, ISTA-Net, is proposed. It is designed to learn optimal parameters, i.e. thresholds, step sizes as well as nonlinear transforms, without hand-crafted settings, in an end-to-end manner. The ISTA-Net method adopts CNN to approximate the nonlinear sparse transformation and improves the recovery performance compared to conventional CS algorithm. As a deep unfolding method for CSI feedback, TiLISTA \cite{Wang20LISTA} utilizes a sparse auto-encoder to learn the sparse transformation in the spatial domain. Nevertheless, due to the truncation of the ISTA algorithm into a finite and fixed number of layers for both training and inference stages, ISTA-Net and TiLISTA struggle with scaling to accommodate a variable number of iterations and face challenges in guaranteeing convergence upon implementation. These problems motivate us to propose a new model-driven DL-based network for CSI feedback with provable convergence guarantee.

\section{Proposed Csi-L2O Method} \label{sec:methodology}
In this section, we propose a new model-driven DL approach, Csi-L2O, which embraces the wisdom of wireless domain knowledge and AI for CSI feedback in FDD massive MIMO systems. We will introduce the general architecture of the proposed Csi-L2O framework, the learnable linear projection at the encoder side, the angular domain sparse transformation at the decoder, and the element-wise L2O decoding module, respectively.

\begin{figure*}[tbh]
\begin{centering}
\subfloat[\label{ChannelsGrey}][COST2100 channel when $N_a = N_t =32$.]{\begin{centering}
\includegraphics[width=0.35\textwidth]{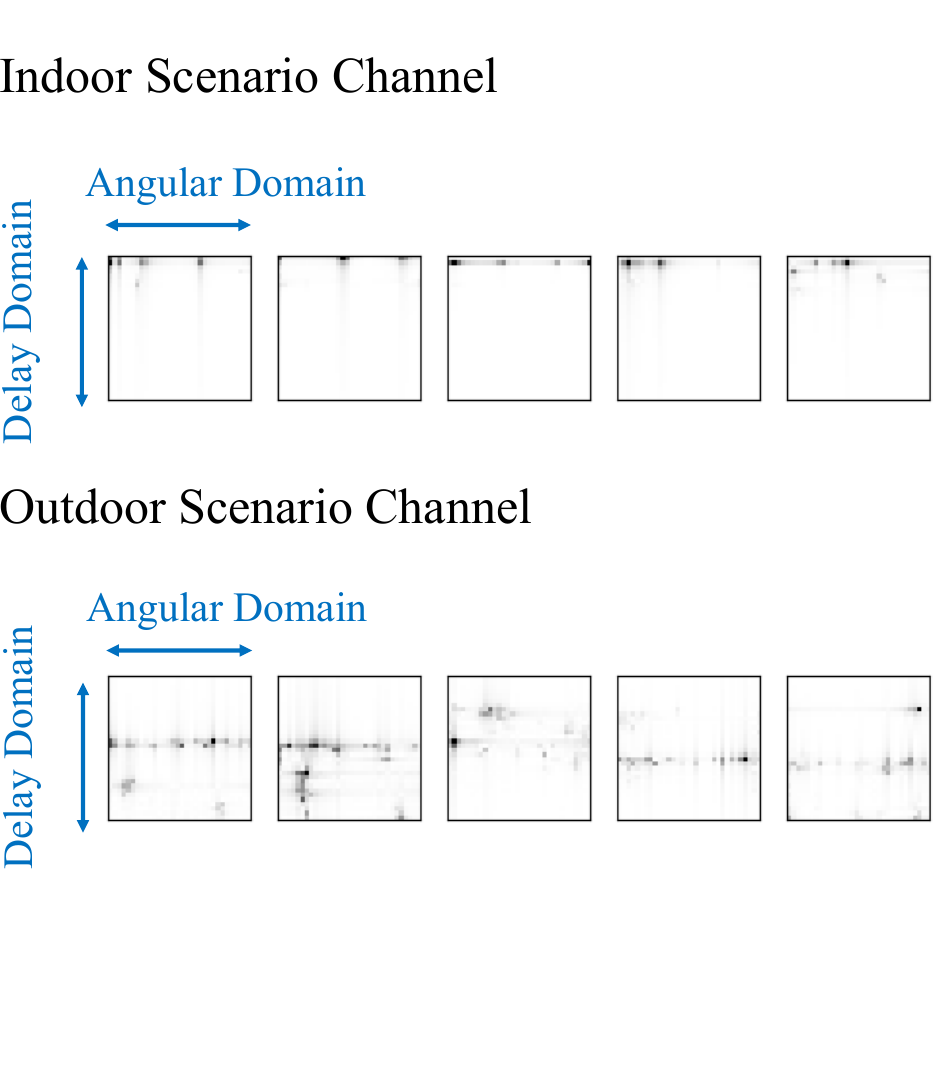}
\par\end{centering}
}\subfloat[\label{ft_fi}][The structure of the proposed angular domain sparse transformation.]{\begin{centering}
\includegraphics[width=0.65\textwidth]{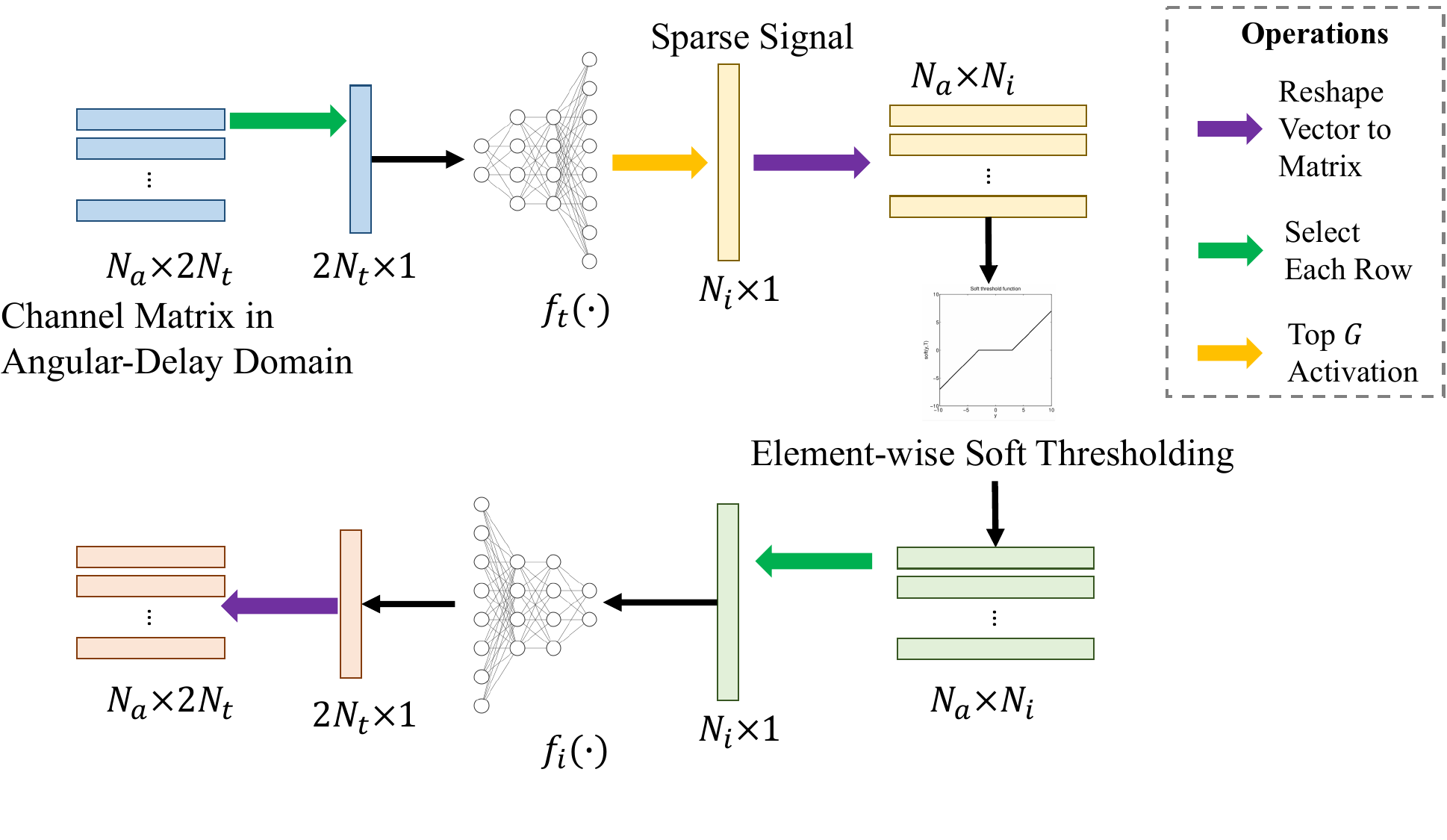}
\par\end{centering}
}\caption{(a) The visualization of indoor and outdoor channels in angular-delay domain generated by COST2100 model when $N_a = N_t =32$. (b) The neural network structure of the proposed angular domain sparse transformation. \label{fig:ft_fi}}
\par\end{centering}
\end{figure*}

\subsection{Architecture of Csi-L2O}
In alignment with the CSI compression and feedback procedure, the proposed Csi-L2O architecture consists of two modules: A compression module and a reconstruction module. The overall architecture of the proposed Csi-L2O is shown in Fig. \ref{Architecture}(a). From the insight of CS, the encoding side is a linear projection and the decoding side is an iterative recovery algorithm. This fits the practical requirement of CSI feedback problem, i.e., the encoder is typically resource-constrained while the decoder enjoys powerful computational capability. At the encoder side, according to Eqn. \eqref{compress}, we employ a learnable projection to compress the CSI where the sampling matrix $\mathbf{W}$ is set learnable and instantiated as a linear layer. Therefore, the encoder is a lightweight and memory-efficient encoding module. 
Concurrently, at the decoding side, the proposal is enhanced with two distinctively engineered components: a learnable sparse transformation and an element-wise L2O mechanism. 
The learnable sparse transformation is designed to identify a sparse representation of CSI in the angular domain, which capitalizes on the inherent sparsity of the channel and consequently improves reconstruction precision. 
Furthermore, different from traditional deep unfolding methods that unroll an existing CS algorithm, we adopt an L2O framework that autonomously learns an optimization algorithm from data. Optimization parameters of $t$-th iteration are colored red in Fig. \ref{Architecture}(a), e.g., preconditioner $\mathbf{p}^{[t]}$, thresholding parameter $\alpha^{[t]}$, accelerator $\mathbf{a}^{[t]}$. These parameters are learned using element-wise L2O module, which is elaborated in Section \ref{Element-wise-L2O}. 
%
Different from existing fully data-driven DL-based methods that treat CSI matrix as a 2D image, the proposed CSI feedback scheme preserves the CS processing pipeline and takes the physical meaning of the wireless channel and sparse recovery into consideration. 

\subsection{Encoder: Learnable Linear Projection}
As shown in \eqref{compress}, traditional CS infers the original signal $\mathbf{h_\text{vec}}$ from the randomized CS measurements $\mathbf{s}$, where $\mathbf{W}$ is a linear random projection matrix. It is important to emphasize that the design of the sampling matrix $\mathbf{W}$ plays a crucial role in preserving the essential elements of the original signal. Researchers have devoted large efforts for developing optimal sampling matrices that contain as much information from the original signals as possible \cite{candes2006stable}. Three types of sampling matrices were proposed in the CS context, which are random, deterministic, and partially orthogonal sampling matrices \cite{bah2010improved}.

In this paper, by capitalizing on the powerful representation ability of DL, we make the matrix $\mathbf{W}$ learnable. The sampling process at the encoder is efficiently implemented as a simple linear layer neural network. $\mathbf{W}$ is naturally the learnable weight of a single fully-connected layer without bias. The sampling matrix is thus able to be trained end-to-end with the decoding module, enabling a good reconstruction accuracy and low encoder-side complexity. Different from conventional fully data-driven method which typically adopts convolutional kernels, fully connected layers, or attention mechanism, our encoder design requires lower computational and memory cost, and thus is more sutable for practical resource-constrained devices.

\subsection{Decoder: Angular Domain Sparse Transformation}
At the decoder side, after receiving the codeword $\mathbf{x}$, the channel reconstruction problem is formulated as
\begin{equation} \label{reconstruction}
\min_{\mathbf{x}}  \frac{1}{2} ||\mathbf{s} - \mathbf{Wx}||_2^2 + \lambda ||f_t(\mathbf{x})||_1,
\end{equation}
where $f_t(\cdot)$ denotes the sparse transformation and $\lambda$ is the regularization parameter. 
While sparse transformation is widely utilized in numerous signal compression methodologies, identifying a transformation basis that can sufficiently sparsify CSI remains a challenging task.

\subsubsection{Channel Sparsity Observations}
Since wireless channels are typically non-stationary, traditional fixed domains, e.g., DFT wavelet transformation, usually result in poor reconstruction performance. In practice, the spatial angles are continuous rather than discrete, which makes the sparsity of the channel coefficients after DFT transformation still insufficient \cite{Wang20LISTA}. 
To demonstrate this conclusion, we plot the gray-scale channel visualizations in angular-delay domain in Fig. \ref{fig:ft_fi}(a). We can observe from Fig. \ref{fig:ft_fi}(a) that due to the multipath effect, there is a high level of sparsity in the delay domain, i.e., only a few elements in each column of channel matrx $\mathbf{H}''$ contains significant values. However, in angular domain (each row of the channel matrix), the sparsity level is still unsatisfactory. This reveals that the signals after the DFT transformation is still not strictly sparse when the number of antennas is not large enough. Besides, due to the complicated outdoor communication surroundings, the sparsity level of outdoor scenario is less satisfactory compared to the indoor scenario.
In this paper, considering the characteristic of wireless channels, we design a learnable angular domain sparse transformation for CSI feedback. 

\subsubsection{Neural Network Design}
The details of the sparse transformation and the inverse transformation are shown in Fig. \ref{fig:ft_fi}(b). To enhance the sparsity level in the angular domain, each row of the channel matrix $\mathbf{H}''$ is selected and fed into the neural network individually. We employ an MLP with three fully-connected layers as $f_t(\cdot)$. $N_i$ denotes the output dimension of $f_t(\cdot)$. In order to obtain strictly sparse signals, only the largest $G$ values of the output of $f_t(\cdot)$ are retained and all the other values are set zero, referring to the top $G$ activation. By doing this, the proposed learning-based sparse transformation function transforms angular domain channels into another domain with sparse features. The inverse transformation $f_i(\cdot)$ exhibits a reverse structure compared to $f_t(\cdot)$. It maps the channels in the learned sparse domain back to the angular domain. The rows of $\mathbf{H}''$ are processed in parallel. After obtaining the output of $f_i(\cdot)$, the estimated channel matrix can be constructed by stacking rows into a whole matrix.

The proposed $f_t(\cdot)$ and $f_i(\cdot)$ guarantee sparsity in the transformed signals and strive to ensure that the signals, when inversely transformed, closely resemble the original ones. $f_t(\cdot)$ and $f_i(\cdot)$ are trained end-to-end with other learning components and the training loss is
\begin{equation} \label{loss}
\begin{aligned}
\text{Loss} = \frac{1}{D} \sum_{i = 1}^{D} & \{ ||\mathbf{H}''_i - \mathcal{D}_{\theta_\mathrm{d}}(\mathcal{E}_{\theta_\mathrm{e}}(\mathbf{H}''_i))||_2^2 + \\
&\beta ||\mathbf{H}''_i - f_i(f_t(\mathbf{H}''_i))||_2^2 \},
\end{aligned}
\end{equation}
where $\mathbf{H}''_i$ denotes the $i$-th channel matrix in the traning dataset, $D$ denotes the total number of training samples, and $\beta$ denotes the balancing term between channel recovery MSE and the sparse transformation MSE.
The proposed sparse transformation effectively overcomes the shortcomings of manually designed transformations for wireless channels. It seeks to discover a sparse basis specifically within the angular domain of the CSI matrix. 
Moreover, the sparse transformation and inverse transformation learned from the numerous CSI training data is more consistent with the data of the specific channel model \cite{Wang20LISTA}. Therefore, the learnable sparse transformation can obtain a more effective sparse representation of CSI, which improves the reconstruction accuracy of the proposed network.


\subsection{Decoder: Element-Wise L2O} \label{Element-wise-L2O}

In order to tackle Problem \eqref{reconstruction}, we propose the L2O strategy that entails parameterizing the update rule into a learnable model. Different from existing CS method that adopts a tedious hand-crafted iterative recovery algorithm, we propose an autonomous learned optimization algorithm from data. 

\subsubsection{Proposed L2O Structure}
Let $F(\mathbf{x})$ denote the objective function of \eqref{reconstruction}. Conventional CS algorithms, e.g., ISTA, solve Problem \eqref{reconstruction} via proximal gradient descent. However, the use of fixed update rule and manually designed optimization parameters leads to a large number of iterations and high computational cost. In contrast to ISTA, we propose to learn the update rule from data to boost decoder-side convergence. The proposed method is designed to determine the update directions by taking the current estimate, i.e.,  $\mathbf{x}^{[t]}$, and the gradient of the objective function, i.e., $\nabla F(\mathbf{x}^{[t]})$, as inputs. The general update rule of the $t$-th iteration is written as:
\begin{equation}
    \label{eq:l2o-general}
    \mathbf{x}^{[t+1]} = \mathbf{x}^{[t]} - \mathbf{d}^{[t]}(\mathbf{z}^{[t]})
\end{equation}
where $\mathbf{d}^{[t]}: \mathcal{Z}\to\mathbb{R}^{2N_a N_t}$ denotes the update direction, $\mathbf{z}^{[t]}\in\mathcal{Z}$ is the input vector, and $\mathcal{Z}$ is the input space. The input vector involves dynamic information, for example $\{\mathbf{x}^{[t]}, F(\mathbf{x}^{[t]}), \nabla F(\mathbf{x}^{[t]})\}$. 
We assume that the update rule $\mathbf{d}^{[t]}(\cdot)$ is differentiable with respect to the input $\mathbf{z}^{[t]}$ and its Jacobian is bounded by a scalar $C$. Formally speaking, the space of update rules is as follows.
\begin{defn}\label{define:d-space}
[Space of Update Rules \cite{Yin23MS4L2O}].
Let $\mathrm{J} \mathbf{d}(\mathbf{z})$ denote the Jacobian matrix of operator $\mathbf{d}:\mathcal{Z}\to\mathbb{R}^{2 N_a N_t}$ and $\|\cdot\|_{\mathrm{F}}$ denote the Frobenius norm, we define the space:
\[
\begin{aligned}
 \mathcal{D}_{C}(\mathcal{Z}) = \Big\{\mathbf{d}: \mathcal{Z} \to \mathbb{R}^{2 N_a N_t} ~\big|~  \mathbf{d} \textnormal{ is differentiable,~~~} \\
\| \mathrm{J}\mathbf{d}(\mathbf{z})\|_{\mathrm{F}} \leq C,~\forall \mathbf{z} \in \mathcal{Z} \Big\}.
\end{aligned}
\]
\end{defn}
In practice, training the deep neural network that is parameterized from $\mathbf{d}^{[t]}(\cdot)$ will require the derivatives of $\mathbf{d}^{[t]}(\cdot)$. Therefore, the differentiablility and bounded Jacobian of the update direction are important. Note that many existing L2O approaches, e.g., LSTM in \cite{andrychowicz2016learning, Lv2017}, achieve $\mathbf{d}^{[t]}(\cdot) \in \mathcal{D}_{C}(\mathcal{Z})$. The definition of the update rule space will help guanrantee the convergence of the proposed L2O scheme, which is proved in Section \ref{convergence-analysis}.

Note that the objective function in Problem \eqref{reconstruction} contains a smooth fidelity function $f(\mathbf{x})=\frac{1}{2} ||\mathbf{s} - \mathbf{Wx}||_2^2$ and a non-smooth regularization function $r(\mathbf{x}) = \lambda ||f_t(\mathbf{x})||_1$. For the smooth part, $\mathbf{x}^{[t]}$ and $\nabla f(\mathbf{x}^{[t]})$ are taken as the input to the update rule. For the non-smooth part, a subgradient $\mathbf{g}^{[t]}$ of $r(\mathbf{x})$ can be utilized. However, the convergence of subgradient descent is generally unstable, and it will not converge to the solution if the step size is  constant \cite{Bertsekas2015}. Proximal Point Algorithm (PPA) \cite{Rockafellar1976} converges faster and more stably than the subgradient descent method. While subgradient descent adopts explicit update, the PPA method takes implicit update rule, i.e., 
\begin{equation}
\mathbf{x}^{[t+1]} = \mathbf{x}^{[t]} - \alpha_\text{PPA}\mathbf{g}^{[t+1]},
\end{equation}
where $\alpha_\text{PPA}$ denotes the step size of PPA algorithm. Inspired by PPA, we select $\mathbf{x}^{[t+1]}$ and $\mathbf{g}^{[t+1]}$ to be the input to the update rule $\mathbf{d}^{[t]}(\cdot)$.

In addition to $\mathbf{x}^{[t]}$, $\nabla f(\mathbf{x}^{[t]})$, $\mathbf{x}^{[t+1]}$, and $\mathbf{g}^{[t+1]}$, we also introduce an auxiliary input $\mathbf{y}^{[t]}$ to $\mathbf{d}^{[t]}(\cdot)$ which contains information about the past estimates and is able to accelerate convergence.
Recall that the update schemes \eqref{ISTA} of existing deep unfolding methods introduced in previous section explicitly depend on only the current status $\mathbf{x}^{[t]}$. Therefore, they lose the ability to capture dynamics between iterations and tend to memorize the datasets. To address this drawback, in the proposed method, we introduce an auxiliary variable $\mathbf{y}^{[t]}$ that encodes historical information through an operator $\mathbf{m}$:
 \begin{equation}
     \label{eq:l2o-nesterov-y}
     \mathbf{y}^{[t]} = \mathbf{m}(\mathbf{x}^{[t]},\mathbf{x}^{[t-1]},\cdots,\mathbf{x}^{[t-K]}),
 \end{equation}
where in addition to the current estimate $\mathbf{x}^{[t]}$, the past $K$ iterations estimates are also taken into consideration.
To facilitate parameterization and training, we assume $\mathbf{m}$ is differentiable, i.e., $\mathbf{m} \in \mathcal{D}_{C}(\mathbb{R}^{ (T+1)\times 2 N_a N_t})$. With the help of $\mathbf{y}^{[t]}$, we are able to infuse more information into the update rule. We set the current estimate, the gradient, the current auxiliary variable, and the gradient of the auxiliary variable as the inputs of the update rule $\mathbf{d}^{[t]}$. The update rule is then shown as \cite{Yin23MS4L2O}
\begin{equation}\label{eq:l2o-nesterov-x}
 \begin{aligned}
\mathbf{x}^{[t+1]}  = \mathbf{x}^{[t]} - \mathbf{d}^{[t]}(&\mathbf{x}^{[t]},\nabla f(\mathbf{x}^{[t]}),\mathbf{x}^{[t+1]},\mathbf{g}^{[t+1]}, \\
&\mathbf{y}^{[t]},\nabla f(\mathbf{y}^{[t]})).
 \end{aligned}
\end{equation}

Follow the derivation in \cite[Theorem 4]{Yin23MS4L2O}, a good update rule should satisfy asymptotic fixed point condition and global convergence condition. We then derive a math-structured update rule from generic update rule \eqref{eq:l2o-nesterov-x}, i.e., for any bounded matrix sequence $\{\mathbf{B}^{[t]}\}_{t=1}^{\infty}$, there exist 
\begin{equation}\label{MS4L2O}
\begin{aligned}
      \mathbf{x}^{[t+1]}  & = \mathbf{x}^{[t]} - (\mathbf{P}_{1}^{[t]}-\mathbf{P}_{2}^{[t]})\nabla f(\mathbf{x}^{[t]}) - \mathbf{P}_{2}^{[t]}\nabla f(\mathbf{y}^{[t]}) - \mathbf{b}_{1}^{[t]} \\ 
      &- \mathbf{P}_{1}^{[t]} \mathbf{g}^{[t+1]} + \mathbf{B}^{[t]}(\mathbf{y}^{[t]} - \mathbf{x}^{[t]}), \\
  \mathbf{y}^{[t+1]} & = (\mathbf{I} - \mathbf{A}^{[t]})\mathbf{x}^{[t+1]} + \mathbf{A}^{[t]}\mathbf{x}^{[t]} + \mathbf{b}_{2}^{[t]},
\end{aligned}
\end{equation}
for all $t=1,2,\cdots$, with $\{\mathbf{P}_{1}^{[t]},\mathbf{P}_{2}^{[t]},\mathbf{A}^{[t]}\}$ being bounded, and $\mathbf{b}_{1}^{[t]}\to \mathbf{0}, \mathbf{b}_{2}^{[t]}\to \mathbf{0}$ as $t \to \infty$. If we further assume $\mathbf{P}_{1}^{[t]}$ is uniformly symmetric positive definite, then we can substitute $\mathbf{P}_{2}^{[t]} {\mathbf{P}_{1}^{[t]}}^{-1}$ with $\mathbf{B}^{[t]}$ and obtain
\begin{equation}
    \label{MS4L2O_2}
    \begin{aligned}
    \hat{\mathbf{x}}^{[t]} &= \mathbf{x}^{[t]} - \mathbf{P}_{1}^{[t]}\nabla f(\mathbf{x}^{[t]}),\\
    \hat{\mathbf{y}}^{[t]} &= \mathbf{y}^{[t]} - \mathbf{P}_{1}^{[t]}\nabla f(\mathbf{y}^{[t]}),\\
     \mathbf{x}^{[t+1]} &= \operatorname{prox}_{r, \mathbf{P}_{1}^{[t]}} \Big( (\mathbf{I} - \mathbf{B}^{[t]})\hat{\mathbf{x}}^{[t]} + \mathbf{B}^{[t]} \hat{\mathbf{y}}^{[t]} - \mathbf{b}_{1}^{[t]} \Big),\\
     \mathbf{y}^{[t+1]} &= \mathbf{x}^{[t+1]} + \mathbf{A}^{[t]} (\mathbf{x}^{[t+1]} - \mathbf{x}^{[t]}) + \mathbf{b}_{2}^{[t]},
    \end{aligned}
\end{equation}
where $\operatorname{prox}_{r, \mathbf{P}_{1}^{[t]}}(\cdot)$ denotes the proximal operator and is defined as
\begin{equation}
    \label{eq:define-prox}
    \operatorname{prox}_{r, \mathbf{P}}(\Bar{\mathbf{x}}):= \argmin_{\mathbf{x}} r(\mathbf{x}) + \frac{1}{2}\|\mathbf{x} - \Bar{\mathbf{x}} \|^2_{\mathbf{P}^{-1}}.
\end{equation}
The norm $\|\cdot\|_{\mathbf{P}^{-1}}$ is defined as $\|\mathbf{x}\|_{\mathbf{P}^{-1}} := \sqrt{\mathbf{x}^\top \mathbf{P}^{-1} \mathbf{x}}$.

In the update scheme \eqref{MS4L2O_2}, $\mathbf{b}_{1}^{[t]}$ and $\mathbf{b}_{2}^{[t]}$ are biases; $\mathbf{A}^{[t]}$ is an accelerator term which can be viewed as an extension of Nesterov momentum; $\mathbf{P}_{1}^{[t]}$ is the preconditioner that plays a similar role as step size in the gradient descent; $\mathbf{B}^{[t]}$ is a balancing term between $\hat{\mathbf{x}}^{[t]}$ and $\hat{\mathbf{y}}^{[t]}$. If $\mathbf{B}^{[t]}=\mathbf{0}$, then $\mathbf{x}^{[t+1]}$ only depends on $\mathbf{x}^{[t]}$ and if $\mathbf{B}^{[t]}=\mathbf{1}$, then $\mathbf{x}^{[t+1]}$ only depends on $\mathbf{y}^{[t]}$ explicitly. Note that ISTA is a special case of update rule \eqref{MS4L2O_2}. When $\mathbf{B}^{[t]} = \mathbf{A}^{[t]} = \mathbf{b}_{1}^{[t]} = \mathbf{b}_{2}^{[t]} = \mathbf{0}$, \eqref{MS4L2O_2} reduces to ISTA. 
Therefore, \eqref{MS4L2O_2} provides more degrees of freedom and is able to enhance reconstruction performance.

To obtain a better balance between performance and efficiency, in our Csi-L2O decoding module, $\mathbf{P}_{1}^{[t]}$, $\mathbf{B}^{[t]}$, and $\mathbf{A}^{[t]}$ are implemented as diagonal matrices, i.e.,
\[\mathbf{P}_{1}^{[t]} = \mathrm{diag}(\mathbf{p}^{[t]}),~~\mathbf{B}^{[t]} = \mathrm{diag}(\mathbf{b}^{[t]}),~~\mathbf{A}^{[t]} = \mathrm{diag}(\mathbf{a}^{[t]}),\]where $\mathbf{p}^{[t]},\mathbf{b}^{[t]},\mathbf{a}^{[t]} \in \mathbb{R}^{2 N_a N_t \times 1}$. The proximal operator is set a scaled soft-thresholding operator, which is expressed as
\begin{equation}
\operatorname{prox}_{\theta^{[t]}}(\mathbf{x}^{[t]}) = \operatorname{sign}\left(\mathbf{x}^{[t]}\right) \max \left(\mathbf{0},\left|\mathbf{x}^{[t]}\right|-\mathbf{\theta}^{[t]}\right), \nonumber
\end{equation}
where $\theta^{[t]}$ denotes the soft-thresholding parameter in the $t$-th iteration.
Update rule \eqref{MS4L2O_2} then becomes:
\begin{equation}
    \label{eq:final-scheme}
     \begin{aligned}
     \hat{\mathbf{x}}^{[t]} &= \mathbf{x}^{[t]} - \mathbf{p}^{[t]}\odot\nabla f(\mathbf{x}^{[t]}),\\
     \hat{\mathbf{y}}^{[t]} &= \mathbf{y}^{[t]} - \mathbf{p}^{[t]}\odot\nabla f(\mathbf{y}^{[t]}),\\
      \mathbf{x}^{[t+1]} &= \operatorname{prox}_{\theta^{[t]}} \Big( (\mathbf{1} - \mathbf{b}^{[t]}) \odot \hat{\mathbf{x}}^{[t]} + \mathbf{b}^{[t]} \odot \hat{\mathbf{y}}^{[t]} - \mathbf{b}_{1}^{[t]} \Big),\\
      \mathbf{y}^{[t+1]} &= \mathbf{x}^{[t+1]} + \mathbf{a}^{[t]} \odot (\mathbf{x}^{[t+1]} - \mathbf{x}^{[t]}) + \mathbf{b}_{2}^{[t]}.
    \end{aligned}
\end{equation}


\begin{table*}[t]	
\selectfont  
\centering
\newcommand{\tabincell}[2]{\begin{tabular}{@{}#1@{}}#2\end{tabular}}
\caption{The computational complexity of different methods.} 
\begin{tabular}{|c|c|c|c|c|}
\hline
Methods & Csi-L2O & CsiNet & TransNet & Deep Unfolding  \\ \hline
Encoder Complexity & $O(N_a N_t M)$ & $O(N_a N_t K_\text{en}^2 C_\text{in} C_\text{out} + N_a N_t M)$ & $O(2(2 N_a^2 d_\text{en} + \frac{1}{2}N_a d_\text{en}^2$ & $O(N_a N_t M)$\\
&&& $+ d_\text{en} N_a N_t) + N_a N_t M)$ &  \\ \hline
Decoder Complexity & $O(T_\text{L2O}(2C_{f_i}+ C_\text{LSTM}))$ & $O( 2 (\sum_{i=1}^{3} N_a N_t K_\text{de,i}^2 C_\text{in,i} C_\text{out,i})$ & $O(2 (4 N_a^2 d_\text{de} + N_a d_\text{de}^2$ & $O(T_\text{DU}(2C_{ST}+ C_\text{ISTA}))$ \\ 
&& $ + N_a N_t M) $& $+ d_\text{de} N_a N_t) + 2 N_a N_t M)$ &  \\ \hline
\end{tabular}
\label{Complexity-BigO}
\end{table*}

\subsubsection{Neural Network Design}
To generate the most appropriate decoding algorithm, the optimization parameters $\mathbf{p}^{[t]}$, $\mathbf{a}^{[t]}$, $\mathbf{b}^{[t]}$, $\mathbf{b}_{1}^{[t]}$, and $\mathbf{b}_{2}^{[t]}$ are not selected mannually but learned from a large amount of data. Note that $\mathbf{p}^{[t]}, \mathbf{a}^{[t]}, \mathbf{b}^{[t]}, \mathbf{b}_{1}^{[t]}, \mathbf{b}_{2}^{[t]} \in \mathbb{R}^{2N_a N_t \times 1}$. In FDD massive MIMO systems, $N_a$ and $N_t$ are large. If a black-box neural network is adopted to model these optimization parameters, the training of the giant and unstructured neural network will be very difficult. In addition, for FDD massive MIMO CSI feedback, the compression ratio needs to be adjusted according to the dynamic communication environment. A reconstruction algorithm that enjoys good generalization ability is thus greatly in need. By taking these two aspects into consideration, we design an element-wise L2O mechanism. In contrast to traditional deep unfolding methods that directly set optimization parameters $\mathbf{p}^{[t]}$, $\mathbf{a}^{[t]}$, $\mathbf{b}^{[t]}$, $\mathbf{b}_{1}^{[t]}$, and $\mathbf{b}_{2}^{[t]}$ trainable, we model them as the output of an element-wise LSTM that greatly improves the generalization ability to different problem scale. The element-wise LSTM is parameterized by learnable parameters $\phi_\text{LSTM}$ and takes the current estimate $\mathbf{x}^{[t]}$ and the gradient $\nabla f(\mathbf{x}^{[t]})$ as the input:
\begin{equation}
    \label{eq:final-scheme-lstm}
    \begin{aligned}
     \mathbf{c}^{[t]},\mathbf{e}^{[t]} = \mathrm{LSTM}\big(\mathbf{x}^{[t]},\nabla f(\mathbf{x}^{[t]}),\mathbf{e}^{[t-1]}; \phi_\text{LSTM} \big),&\\
     \mathbf{p}^{[t]}, \mathbf{a}^{[t]}, \mathbf{b}^{[t]}, \mathbf{b}_{1}^{[t]}, \mathbf{b}_{2}^{[t]}= \mathrm{MLP}(\mathbf{c}^{[t]}; \phi_{\text{MLP}}),&\\
    \end{aligned}
\end{equation}
where $\mathbf{e}^{[t]}$ is the internal state of LSTM, $\mathbf{e}^{[0]}$ is randomly sampled from Gaussian distribution, and $\mathbf{c}^{[t]}$ is the output of LSTM which is then fed into the MLP to generate the optimization parameters. Detailed procedure is demonstrated in Fig. \ref{Architecture}(b). 
An ``element-wise'' LSTM means that the same network is shared across all coordinates of the input. Specifically, each coordinate of $\mathbf{x}^{[t]}$ and $\nabla f(\mathbf{x}^{[t]})$ are fed into the LSTM in parallel. With this method, the single model can be applied to optimization problems of any scale and thus fits the variable compression ratio cases.
It is common in classic optimization algorithms to take positive $\mathbf{p}^{[t]}$ and $\mathbf{a}^{[t]}$. Therefore, we use an additional activation function to post-process $\mathbf{p}^{[t]}$ and $\mathbf{a}^{[t]}$, e.g., sigmoid function.
\eqref{eq:final-scheme} and \eqref{eq:final-scheme-lstm} together define the L2O decoding scheme.

\subsubsection{Comparison with Deep Unfolding}
Key differences between Csi-L2O method and deep unfolding methods include the way of parameterization and the existence of a convergence guarantee. 
On the one hand, different from the element-wise LSTM parameterization \eqref{eq:final-scheme-lstm}, deep unfolding methods make optimization parameters learnable and directly optimize them from data. For example, instead of using neural network to generate $\mathbf{p}^{[t]}, \mathbf{a}^{[t]}, \mathbf{b}^{[t]}, \mathbf{b}_{1}^{[t]}, \mathbf{b}_{2}^{[t]}$, one can directly turn the step size and soft-threshold parameters trainable.
However, this direct parameterization introduces several limitations. It hampers the model's ability to capture dynamics between iterations and leads to a tendency to memorize specific datasets rather than generalizing. Additionally, direct parameterization requires that the dimensions of the learnable parameters match the scale of the problem, which restricts the model's applicability to optimization problems of different scales during inference. This constraint prevents deep unfolding methods from generalizing effectively to various compression ratio cases. 
On the other hand, since deep unfolding algorithms are built by fixed and finite layers, it is difficult to scale to different number of iterations. When the number of layers is different during training and testing, it is hard to ensure convergence of deep unfolding.

\section{Convergence and Complexity Analysis} \label{ComplexityAnalysis}
In this section, we first emphasize the importance and the proof of the convergence for the proposed update rule. Then, the computational complexity analysis of the proposed method and the comparison with other benchmarks are demonstrated.
\subsection{Convergence Analysis}\label{convergence-analysis}
Conventional deep unfolding method typically lacks convergence guarantee, making it difficult to scale to variable number of layers during inference \cite{Wentao23DEQ}. The deployment of different number of layers from training will result in performance fluctuation. In this subsection, we will prove the convergence of the proposed update rule, i.e., $\mathbf{d}^{[t]}(\cdot)$. The convergence guarantee will help us improve the reliablity of the proposed method and determine the appropriate number of layers during inference. 

Let $\mathbf{x}^*$ be the fixed point of Problem \eqref{reconstruction}. We then have the following theorem. 
\begin{thm}\label{thm1} 
For any $\mathbf{x}^* \in \argmin_{\mathbf{x}\in\mathbb{R}^{2N_aN_t}}F(\mathbf{x})$, 
 \begin{align}
   \lim_{t\to \infty} & \mathbf{d}^{[t]}(\mathbf{x}^*,\nabla f(\mathbf{x}^*),\mathbf{x}^*,-\nabla f(\mathbf{x}^*),\mathbf{x}^*,\nabla f(\mathbf{x}^*)) = \mathbf{0}, \label{eq:fp}\\
    & \mathbf{m}(\mathbf{x}^*,\mathbf{x}^*,\cdots,\mathbf{x}^*) = \mathbf{x}^*. \label{eq:fp2}
\end{align}
For any sequences $\{\mathbf{x}^{[t]}, \mathbf{y}^{[t]}\}_{t=0}^{\infty}$ generated by \eqref{eq:l2o-nesterov-y} and \eqref{eq:l2o-nesterov-x}, there exists one $\mathbf{x}^* \in \argmin_{\mathbf{x}\in\mathbb{R}^{2N_aN_t}}F(\mathbf{x})$ such that 
\begin{equation}
    \label{eq:gc4}
\lim_{t\to\infty}\mathbf{x}^{[t]} = \lim_{t\to\infty}\mathbf{y}^{[t]} = \mathbf{x}^*.
\end{equation}
\end{thm}
\begin{IEEEproof}
Please refer to Appendix \ref{proof:thm1}.
\end{IEEEproof}
Eqn. \eqref{eq:fp} shows that the proposed update rule $\mathbf{d}^{[t]}(\cdot)$ guarantees $\mathbf{x}^{[t+1]}=\mathbf{x}^*$ as long as $\mathbf{x}^{[t]} = \mathbf{x}^*$. This means that if 
$\mathbf{x}^{[t]}$ is a solution, the next iteration is also fixed. \eqref{eq:fp} and \eqref{eq:fp2} together guarantee the convergence of the proposed parameterization update rule.

\subsection{Complexity Analysis}
The encoder-side computational complexity of the proposed Csi-L2O and that of other baselines are illustrated in Table \ref{Complexity-BigO}.
Since there is a linear projection at the encoder, the encoder complexity of the proposed Csi-L2O is $O(N_a N_t M)$, which grows linearly with the number of antennas and the dimension of the codeword. The encoder complexity of CsiNet is $O(N_a N_t K_\text{en}^2 C_\text{in} C_\text{out} + N_a N_t M)$, where $K_\text{en}$, $C_\text{in}$, and $C_\text{out}$ denote the dimension of the convolutional kernel, the input and output channel number, respectively. As the encoding module of CsiNet consists of both convolutional kernels and fully connected layers, the computational complexity of the CsiNet is higher than that of the proposed Csi-L2O.
On the other hand, the encoder complexity of TransNet is $O(2(2 N_a^2 d + \frac{1}{2}N_a d_\text{en}^2 + d_\text{en} N_a N_t) + N_a N_t M)$, where $d_\text{en}$ denotes the encoder-side self-attention dimension. The complexity mainly comes from two attention-based encoding blocks and fully connected layers. Although transformer-based autoencoder achieves SOTA performance, it puts prohibitive computational burdens for resource-constrained devices.
The encoder-side complexity of deep unfolding methods, including ISTA-Net and TiLISTA, are both $O(N_a N_t M)$ since they use a linear projection at the encoder.
According to the complexity analysis, there's a guarantee that the proposed method will achieve much higher computational efficiency compared to SOTA method, TransNet. The computational complexity reduction is more obvious when $N_t$ and/or $N_a$ is large, which is indeed the situation that future wirless systems will meet \cite{wang2023tutorial}.

The decoder-side computational complexity of different methods are also shown in Table \ref{Complexity-BigO}. 
The decoder complexity of the proposed Csi-L2O is $O(T_\text{L2O}(2C_{f_i}+ C_\text{LSTM}))$, where $T_\text{L2O}$ denotes the number of layers in the decoder, $C_{f_i}$ is the complexity of the sparse transformation function $f_t(\cdot)$, and $C_\text{LSTM}$ denotes the complexity of LSTM, respectively. The decoder complexity of deep unfolding method exhibits a similar structure, i.e., $O(T_\text{DU}(2C_\text{ST}+ C_\text{ISTA}))$, where $T_\text{DU}$ denotes the number of layers, $C_\text{ST}$ is the complexity of the sparse transformation, and $C_\text{ISTA}$ denotes the complexity of each iteration in ISTA, respectively. 
Besides, the decoder complexity of CsiNet is $O(2 (\sum_{i=1}^{3} N_a N_t K_\text{de,i}^2 C_\text{in,i} C_\text{out,i}) + N_a N_t M)$, where $K_\text{de,i}$, $C_\text{in,i}$, and $C_\text{out,i}$ denote the dimension of the convolutional kernel, the input and output channel number of the $i$-th layer in the CNN, respectively. The decoder complexity of TransNet is $O(2 (4 N_a^2 d_\text{de} + N_a d_\text{de}^2 + d_\text{de} N_a N_t) + 2 N_a N_t M)$, where $d_\text{de}$ denotes the decoder-side self-attention dimension. Although the direct comparison of decoder-side computational complexity among different methods is difficult, we will show the exact values for different approaches via simulations in Section \ref{sec:complexity}. 

\section{Simulation Results}
In this section, we demonstrate the performance of the proposed Csi-L2O network for CSI feedback. We first introduce the dataset generation, training settings, and evaluation metrics. The performance comparison of the proposed approach with several representative baselines are then demonstrated. Next, we discuss the computational complexity and convergence behavior of different DL-based CSI feedback methods. The bit-level performance is also demonstrated, where a quantization module is added to generate zero one bit streams. Finally, the multiple rate feedback scenarios are considered, which validates the superior generalization ability of the proposed Csi-L2O to different compression ratios.
\subsection{Simulation Setup}
\subsubsection{Data Generation}
Following the experimental setting in \cite{CsiNet}, two types of channel matrices are generated according to the COST 2100 models \cite{COST2100}, i.e., the indoor picocellular scenario working at the 5.3 GHz band and the outdoor rural scenario working at the 300 MHz band. The BS is equipped with the uniform linear array with $N_t = 32$ and the number of subcarriers is 1024. The original $2 \times 1024 \times 32$ CSI matrix is transformed into the angular-delay domain and truncated to the first 32 rows, forming the $2 \times 32 \times 32$ matrix $\mathbf{H}''$.
\subsubsection{Training Settings}
The training, validation, and test datasets contain 100,000, 30,000, and 20,000 samples, respectively. The Adam optimizer is used for trainable weight updates \cite{kingma2014adam}. Kaiming initialization is used as the neural network initialization approach. We train the neural network for 1000 epochs with a mini-batch size of 200 and a learning rate of 0.0001. The loss function in \eqref{loss} is used as the unsupervised loss where $\beta$ is set 0.01. $f_t(\cdot)$ is a three-layer MLP with hidden units [128, 128, 256] and $f_i(\cdot)$ exhibits a reverse structure, i.e., a three-layer MLP with hidden units [256, 128, 128]. The top 51 elements are retained in the top $G$ activation of sparse transformation. A two-layer LSTM with hidden size being two is adopted as the element-wise LSTM in the L2O decoding module. A single-layer MLP with 20 input size and 20 output size generates the intermediate parameters, which is then fed into five dstinct single-layer MLPs to output optimization parameters in element-wise L2O. 
\subsubsection{Evaluation Metric}
The normalized mean squared error (NMSE) between the recovered channel and the true channel is used to evaluate the performance, which is given by
\begin{equation}
\text{NMSE} = \mathbb{E} \left\{\frac{||\mathbf{H}'' - \hat{\mathbf{H}}''||_2^2}{||\mathbf{H}''||_2^2}  \right\}.
\end{equation}
In addition, the number of FLOPs is used to measure the time complexity of the learning model, and the number of trainable parameters is adopted as a metric to measure the space complexity \cite{TransNet}. All the simulations are done using the existing DL platform PyTorch. The number of FLOPs and trainable parameters are calculated using the thop package \cite{Thop} for PyTorch.

\subsection{Performance Comparison}
To illustrate the effectiveness of the proposed CSI feedback design, we adopt five benchmarks for comparison:
\begin{itemize}
\item \textbf{ISTA}: A classical CS algorithm without learning component.
\item \textbf{MS4L2O \cite{Yin23MS4L2O}}: A mathematical inspired L2O framework is directly implemeted on CSI feedback problem.
\item \textbf{CsiNet \cite{CsiNet}}: An exploratory fully data-driven CSI feedback scheme that enjoys low time and space complexity.
\item \textbf{TransNet \cite{TransNet}}: A transformer-based method that achieves SOTA performance but induces heavy computational costs.
\item \textbf{TiLISTA \cite{Wang20LISTA}}: An ISTA-based deep unfolding method for CSI feedback where a sparse auto-encoder is utilized to learn the sparse transformation in the spatial domain.
\end{itemize}

\begin{table*}[t]	
\selectfont  
\centering
\newcommand{\tabincell}[2]{\begin{tabular}{@{}#1@{}}#2\end{tabular}}
\caption{The encoder-side FLOPs and trainable parameters number of different methods.} 
\begin{tabular}{|c|c|c|c|c|c|c|c|c|}
\hline
Compression Ratio & \multicolumn{2}{c|}{$1/8$} & \multicolumn{2}{c|}{$1/16$} & \multicolumn{2}{c|}{$1/32$} & \multicolumn{2}{c|}{$1/64$} \\ \hline
Numbers & FLOPs & Params & FLOPs & Params & FLOPs & Params & FLOPs & Params \\ \hline
ISTA & 0.524 M & 0 & 0.262 M & 0 & 0.131 M & 0 & 0.066 M & 0  \\ \hline
MS4L2O & 0.524 M & 0 & 0.262 M & 0 & 0.131 M & 0 & 0.066 M & 0 \\ \hline
CsiNet  & 0.561 M & 0.524 M & 0.299 M & 0.262 M & 0.168 M & 0.131 M & 0.102 M & 0.066 M\\ \hline
TiLISTA & 0.524 M & 0.524 M & 0.262 M & 0.262 M & 0.131 M & 0.131 M & 0.066 M & 0.066 M \\ \hline
TransNet & 17.334 M & 0.789 M & 17.072 M & 0.526 M & 16.941 M & 0.395 M & 16.876 M & 0.330 M \\ \hline
\textbf{Proposed} & 0.524 M & 0.524 M & 0.262 M & 0.262 M & 0.131 M & 0.131 M & 0.066 M & 0.066 M \\ \hline
\end{tabular}
\label{complexity-en}
\end{table*}

\begin{table*}[t]	
\selectfont  
\centering
\newcommand{\tabincell}[2]{\begin{tabular}{@{}#1@{}}#2\end{tabular}}
\caption{The decoder-side FLOPs and trainable parameters number of different methods.} 
\begin{tabular}{|c|c|c|c|c|c|c|c|c|}
\hline
Compression Ratio & \multicolumn{2}{c|}{$1/8$} & \multicolumn{2}{c|}{$1/16$} & \multicolumn{2}{c|}{$1/32$} & \multicolumn{2}{c|}{$1/64$} \\ \hline
Numbers & FLOPs & Params & FLOPs & Params & FLOPs & Params & FLOPs & Params \\ \hline
ISTA & 10.486 M & 0 & 5.243 M & 0 & 2.621 M & 0 & 1.311 M & 0 \\ \hline
MS4L2O & 20.978 M & 0.004 M & 10.492 M & 0.004 M & 5.249 M & 0.004 M & 2.628 M & 0.004 M \\ \hline
CsiNet & 3.809 M & 0.527 M & 3.547 M & 0.265 M & 3.416 M & 0.134 M & 3.351 M & 0.069 M \\ \hline
TiLISTA & 10.813 M & 0.033 M & 5.571 M & 0.033 M & 2.949 M & 0.033 M & 1.638 M & 0.033 M  \\ \hline
TransNet & 17.883 M &  1.315 M & 17.359 M & 0.791 M & 17.097 M & 0.530 M & 16.966 M & 0.398 M  \\ \hline
\textbf{Proposed} & 22.125 M & 0.119 M & 11.639 M & 0.119 M & 6.396 M & 0.119 M & 3.201 M & 0.119 M \\ \hline
\end{tabular}
\label{complexity-de}
\end{table*}

\begin{figure}[t] 
\centering
\includegraphics[height=6.0cm]{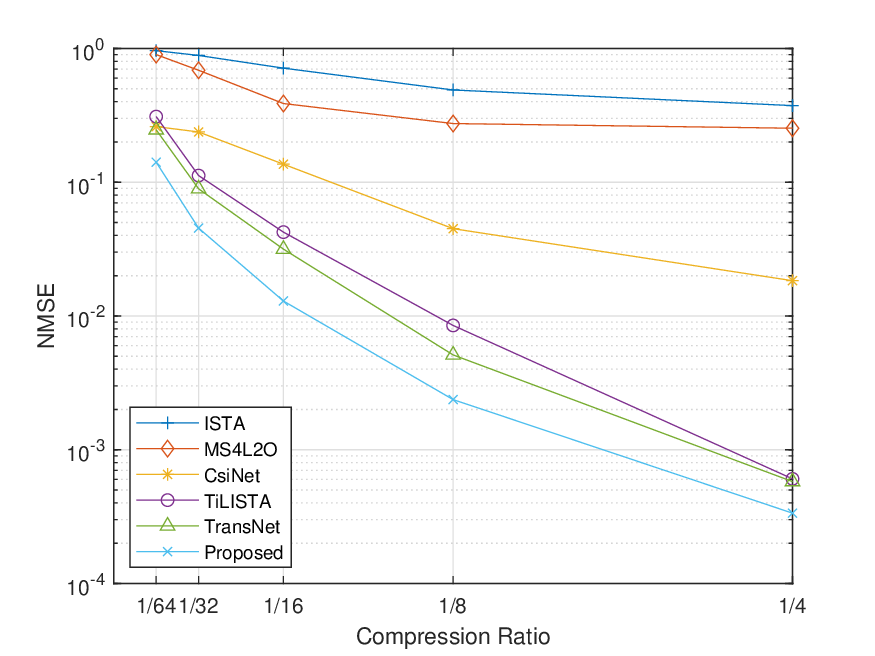} 
\caption{NMSE achieved by different methods versus compression ratios in an indoor scenario.} 
\label{Indoor} 
\end{figure}

Fig. \ref{Indoor} plots the NMSE achieved by the proposed scheme and the five baseline methods versus the compression ratios in indoor scenario. The traditional ISTA performs the worst because the CSI after DFT transformation is not sparse enough. 
It is shown that all the learning-based methods outperform the ISTA method, indicating that DL approaches have the ability to effectively compress and reconstruct CSI. Among the five learning-based methods, the proposed Csi-L2O scheme achieves the best performance for all investigated values of compression ratios. For example, when the compression ratio is $1/16$ the proposed Csi-L2O outperforms SOTA TransNet 3.88 dB. It is also observed that the proposed Csi-L2O design outperforms the MS4L2O to a large margin, and the performance gain is more obvious when the compression ratio is large. This indicates the effectiveness of the proposed learnable sampling matrix at the encoder and the angular domain sparse transformation function at the decoder. 

\begin{figure}[t] 
\centering
\includegraphics[height=6.0cm]{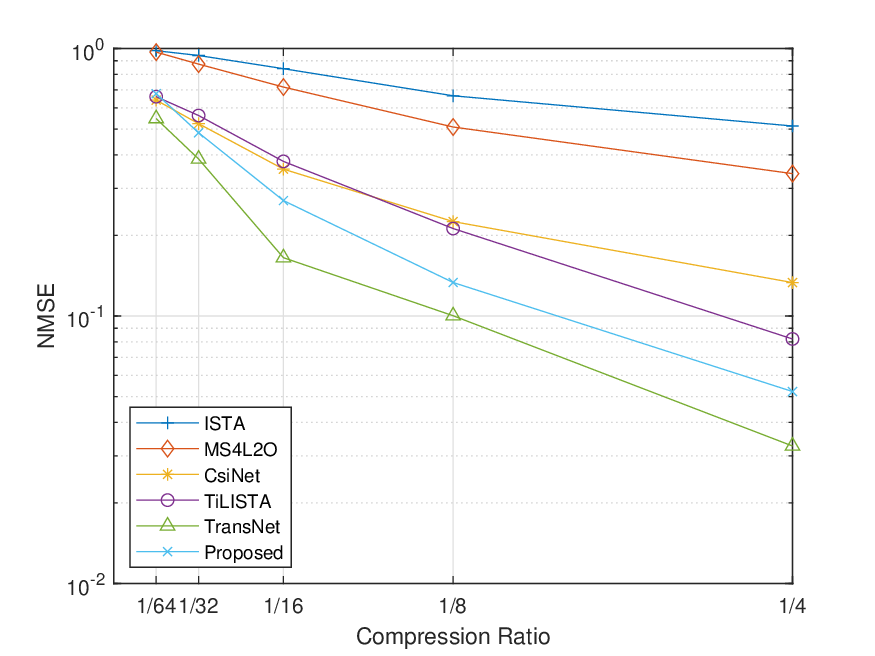} 
\caption{NMSE achieved by different methods versus compression ratios in an outdoor scenario.} 
\label{Outdoor} 
\end{figure}

In Fig. \ref{Outdoor}, we demonstrate the CSI recovery accuracy achieved by different methods versus the compression ratios in outdoor scenario. As can be observed in Fig. \ref{Outdoor}, while the ISTA, MS4L2O, CsiNet, and TiLISTA methods entail a prominent performance loss, our proposed L2O-based method still captures the trend of the SOTA and achieves a comparable performance. This indicates that even for the complicated communication environment, the proposed feedback scheme is still effectively learned from data thanks to the powerful learning capability of LSTM and MLP.

\subsection{Complexity Comparison} \label{sec:complexity}
We then show the number of FLOPs and number of trainable parameters of different methods at the encoder-side in Table \ref{complexity-en} under different compression ratios. Due to two consecutive attention-based encoding blocks and fully connected layers, the TransNet entails the highest time complexity and space complexity, which hinders their applications in practice especially for resource-constraint devices. For example, when the compression ratio is $1/16$, the proposed Csi-L2O only requires $1.5\%$ number of FLOPs of TransNet. Since ISTA, MS4L2O, and TiLISTA all employ a simple linear projection at the encoder, and thus they all enjoy high computational efficiency. ISTA and MS4L2O utilize a Guassian random matrix as the sampling matrix and thus the number of trainable parameters are zeros. In addition to one fully connected layer, CsiNet also adopts convolutional kernels, making the number of FLOPs and number of trainable parameters slightly higher than TiLISTA. It can be observed that the proposed Csi-L2O method has exactly the same encoder-side complexity as TiLISTA, which is the lowest among all the baselines.
Furthermore, the decoder-side computational complexity is demonstrated in Table \ref{complexity-de}. It is observed that the computational complexity of Csi-L2O is less than TransNet when compression ratio is no less than $1/16$. The complexity reduction is more obvious for low compression ratios, indicating the superiority of the proposed method when the number of feedback bits is very limited. Besides, the number of trainable parameters of Csi-L2O is lower than TransNet both at the encoder and decoder, resulting in less memory cost. 

\begin{table*}[t]	
\selectfont  
\centering
\newcommand{\tabincell}[2]{\begin{tabular}{@{}#1@{}}#2\end{tabular}}
\caption{The bit-level CSI feedback NMSE (in dB) of different methods in an indoor scenario.} 
\begin{tabular}{|c|c|c|c|c|c|c|c|c|c|c|}
\hline
Compression Ratio & \multicolumn{5}{c|}{$1/4$} & \multicolumn{5}{c|}{$1/8$}  \\ \hline
Quantization Level & No Quant & B = 3 & B = 4 & B = 5 & B = 6 & No Quant & B = 3 & B = 4 & B = 5 & B = 6 \\ \hline
ISTA & -4.27 & -1.39  & -1.71  & -2.39  & -2.98 & -3.10 & -0.89 & -1.01  & -1.43  & -2.07  \\ \hline
MS4L2O & -5.96 & -2.17 & -2.89 & -3.77 & -4.05 & -5.61 & -2.03 & -2.55 & -3.42 & -3.91 \\ \hline
CsiNet & -17.36 & -9.89 & -11.97 & -13.25 & -14.63 & -13.47 & -5.15 & -6.03 & -8.49 & -10.31 \\ \hline
TransNet & -32.38 & -19.32 & -23.51 & -27.00 & -28.97 & -22.91 & -10.17 & -12.88 & -15.64 & -19.04  \\ \hline
TiLISTA & -32.18 & -18.73 & -22.09 & -26.31 & -28.08 & -20.71 & -9.05 & -11.63 & -14.47 & -17.10 \\ \hline
\textbf{Proposed} & \textbf{-34.74} & \textbf{-21.77} & \textbf{-24.96} & \textbf{-27.83} & \textbf{-30.97} & \textbf{-26.25} & \textbf{-15.41} & \textbf{-17.94} & \textbf{-20.88} & \textbf{-23.85}  \\ \hline
Compression Ratio & \multicolumn{5}{c|}{$1/16$} & \multicolumn{5}{c|}{$1/32$}  \\ \hline
Quantization Level & No Quant & B = 3 & B = 4 & B = 5 & B = 6 & No Quant & B = 3 & B = 4 & B = 5 & B = 6 \\ \hline
ISTA & -1.47  & -0.40 & -0.68  & -0.94  & -1.25 & -0.52 & -0.28 & -0.33 & -0.41 & -0.49 \\ \hline
MS4L2O & -4.11 & -1.05 & -1.97 & -3.03 & -3.91 & -1.62 & -0.58 & -0.71 & -1.14 & -1.50 \\ \hline
CsiNet & -8.65 & -4.65 & -5.70 & -6.91 & -8.43 & -6.24 & -4.02 & -5.31 & -5.78 & -6.12 \\ \hline
TransNet & -15.00 & -9.70 & -11.86 & -14.10 & -14.87 & -10.49 & -8.89 & -9.48 & -9.92 & -10.26  \\ \hline
TiLISTA & -13.73 & -9.11 & -11.02 & -13.31 & -14.01 & -9.50 & -7.73 & -8.11 & -8.71 & -9.08 \\ \hline
\textbf{Proposed} & \textbf{-18.88} & \textbf{-13.94} & \textbf{-15.98} & \textbf{-17.79} & \textbf{-18.61} & \textbf{-13.43} & \textbf{-11.90} & \textbf{-12.40} & \textbf{-12.99} & \textbf{-13.31}  \\ \hline
\end{tabular}
\label{BitLevel}
\end{table*}

\begin{figure}[t] 
\centering
\includegraphics[height=6.0cm]{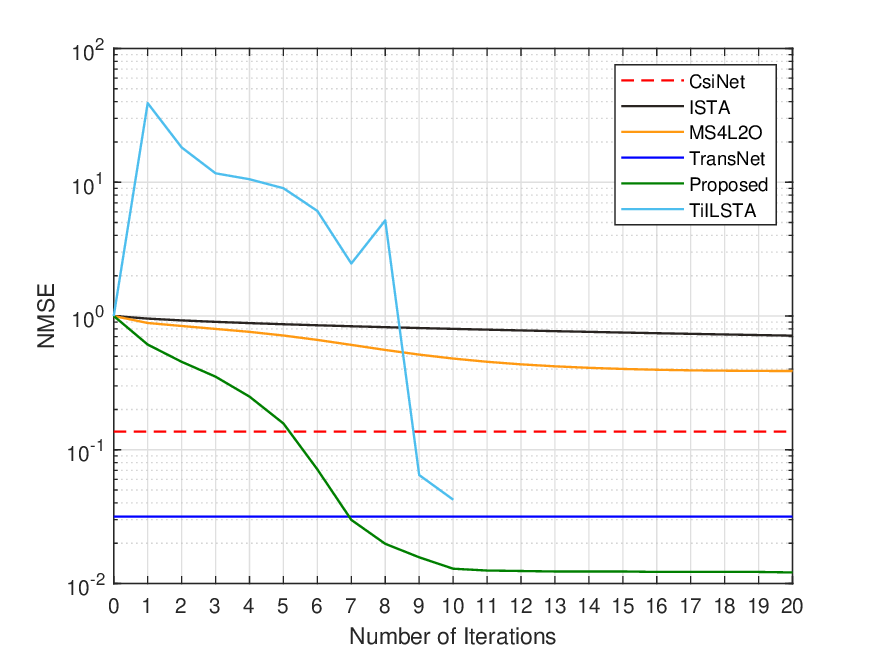} 
\caption{NMSE achieved by different methods versus the number of iterations in an indoor scenario when compression ratio is $1/16$.} 
\label{convergence} 
\end{figure}

\subsection{Convergence}
Fig. \ref{convergence} illustrates the performance comparison among several methods with different numbers of iterations. It is demonstrated that although the ISTA and MS4L2O methods converge quickly, the reconstruction accuracy is still poor. In comparison, the proposed method has significant performance gain in terms of accuracy. Since the fully data-driven baselines, i.e., CsiNet and TransNet, use explicit neural networks and the outputs are acquired through one forward propagation, they do not have the concept of iterations. It can be observed from Fig. \ref{convergence} that the proposed Csi-L2O converges within 11 iterations and running the proposed method with 7 iterations outperforms TransNet. It is also shown that although TiLISTA achieves a comparable performance as TransNet in 10 iterations, it does not guarantee to converge and fluctuates severely. When TiLISTA is trained for 20 iterations, the final NMSE, i.e., $-10.89$ dB, is even worse than that of 10 iterations, i.e., $-13.73$ dB, because deeper deep unfolding algorithm is harder to be trained. Therefore, we plot the convergence curve for 10-iteration TiLISTA. 

\subsection{Bit Level Quantization}
In this subsection, we compare the reconstruction accuracy of different methods in bit level CSI feedback. When the encoding and decoding modules are fixed, in practice, the quantization module is introduced to quantize the compressed codeword into zero one bit streams \cite{Tong19Quantize}. In Table \ref{BitLevel}, we compare the bit-level CSI feedback performance of different methods under different compression ratios in indoor scenario. Non-uniform Lloyd-Max quantizer is adopted as the quantization module \cite{LloydMax}.
In Table \ref{BitLevel}, $B$ denotes the number of quantization bits.
As we can observe, the reconstruction accuracy increases with the increase of quantization bits. Particularly, Csi-L2O with $B=6$ even exhibits a similar performance as the original Csi-L2O without quantization. When the compression ratio is low, e.g., compression ratio is $1/32$, the performance loss due to the quantization is marginal. 

In practical scenarios, the compression ratio and quantization bits $B$ together determine the overhead of CSI feedback. For example, if the feedback bitstream contains 1536 bits, we can have two choices, i.e., compression ratio is $1/4$ and the number of quantization bits is 3, or compression ratio is $1/8$ and the number of quantization bits is 6. The NMSE of the former at the indoor scenario is $-21.77$ dB, while that of the latter is $-23.85$ dB. 
This provides a guidance for the practical deployment that, even if the length of feedback bitstream is fixed, suitable compression ratio and quantization bits have to be selected jointly to achieve the optimal performance.

\begin{figure}[t] 
\centering
\includegraphics[height=6.5cm]{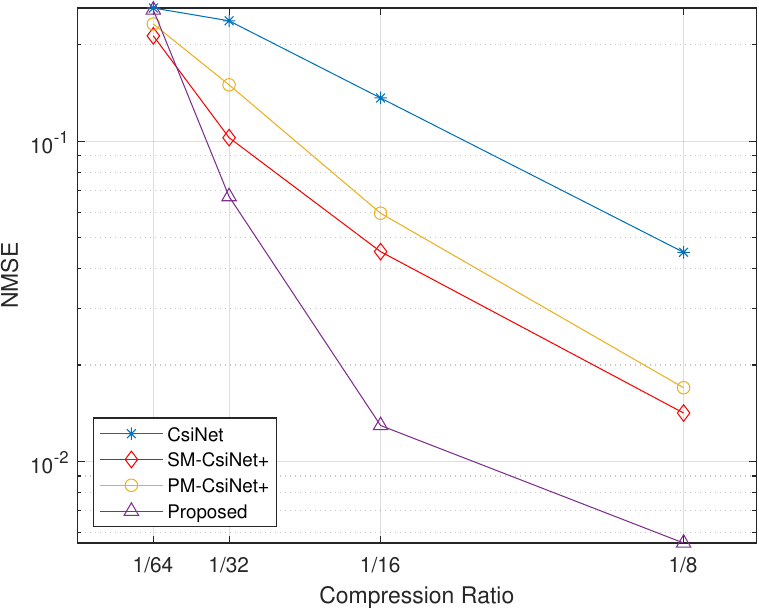} 
\caption{NMSE achieved by different methods versus compression ratios in an indoor scenario.} 
\label{OoDIndoor} 
\end{figure}

\subsection{Multiple-Rate Feedback Scenarios}
In practice, the compression ratio has to be adjusted according to the dynamic environments and varying coherence time \cite{Guo20MultiRate}, named multiple-rate CSI feedback. Fig. \ref{OoDIndoor} shows the NMSE performance of the proposed method for multiple-rate CSI feedback. The proposed method in Fig. \ref{OoDIndoor} is trained when compression ratio is $1/16$ and directly test for other settings. The CsiNet in  Fig. \ref{OoDIndoor} is retrained each time the compression ratio changes. Two baselines specially designed for multiple feedback rate cases are also compared, i.e., SM-CsiNet+ and PM-CsiNet+ \cite{Guo20MultiRate}. SM-CsiNet+ is a serial manner multi-rate CSI feedback method where different compression ratios share the first a few layers of neural network, and the output of high compression ratio part is the input of low compression ratio part. PM-CsiNet+ is a parallel manner multi-rate CSI feedback method where the output of low compression ratio is a part of the output of high compression ratio. 
It is demonstrated in Fig. \ref{OoDIndoor} that the proposed Csi-L2O achieves the best multi-rate feedback reconstruction accuracy among all the baselines when compression ratios are above $1/64$. This verifies that the proposal has good generalization ability. Once trained, the proposed Csi-L2O can be directly implemented to different compression ratios without additional training. 

\begin{table}[t]	
\selectfont  
\centering
\newcommand{\tabincell}[2]{\begin{tabular}{@{}#1@{}}#2\end{tabular}}
\caption{The encoder-side FLOPs and trainable parameters number of different methods in multiple feedback rate scenarios.} 
\begin{tabular}{|c|c|c|}
\hline
Methods & Number of FLOPs & Number of Trainable Parameters  \\ \hline
SM-CsiNet+ & 1.638 M & 1.222 M   \\ \hline
PM-CsiNet+ & 1.466 M & 1.649 M    \\ \hline
Proposed & 0.262 M &  0.262 M    \\ \hline
\end{tabular}
\label{OoD-complexity}
\end{table}

Table \ref{OoD-complexity} then compares the encoder-side computational complexity of the proposed method with SM-CsiNet+ and PM-CsiNet+. Note that in multiple rate feedback case, all the considered three methods have fixed complexity, i.e., the complexity is independent on the compression ratios. It is demonstrated in Table \ref{OoD-complexity} that since SM-CsiNet+ adopts the deepest neural network for compression, it has the highest time complexity. The number of FLOPs of PM-CsiNet+ is lower than that of SM-CsiNet+ because of the layer reuse in the parallel structure. The proposed method achieves the lowest number of FLOPs, nearly $16\%$ of SM-CsiNet+, showing the high computational efficiency. For the space complexity, since the proposed Csi-L2O only adopts a simple linear layer, it has the least number of trainable parameters.

\section{Conclusions}
In this paper, we developed a model-driven DL-based method, Csi-L2O, for CSI feedback in FDD massive MIMO systems. In contrast to the existing DL-based CSI feedback paradigm, i.e., fully data-driven methods, we proposed an innovative way to amalgamate domain knowledge with DL. In particular, the codeword is generated via a learnable linear projection at the user side, while the full CSI is reconstructed at the BS side using an element-wise parameterized update rule. The proposal features an encoder with extremely low complexity, offers performance that rivals SOTA solutions, and has the flexibility to adjust to multiple feedback rates without necessitating the retraining of the neural network. Simulation results clearly demonstrated that the proposed Csi-L2O achieves an excellent performance. It is intriguing to extend our proposed Csi-L2O to other challenging communication applications, such as multi-cells massive MIMO systems \cite{Ma22DeepUnfolding}, CSI feedback in movable antenna systems \cite{xiao2023channel}, and CSI feedback with time variant channels \cite{Markovian22}.
\appendices 
\section{Proof of Theorem \ref{thm1}} \label{proof:thm1}
\begin{IEEEproof}
Before the proofs of \textbf{Theorem \ref{thm1}}, we first introduce a lemma proved in \cite[Lemma 1]{Yin23MS4L2O} to facilitate our proof.

\begin{lem}\label{lemma:mvt}
For any operator $\mathbf{o} \in \mathcal{D}_{C}(\mathbb{R}^{m\times n})$ and any $\mathbf{x}^{[1]},\mathbf{y}^{[1]},\mathbf{x}^{[2]},\mathbf{y}^{[2]},\cdots,\mathbf{x}^{[m]},\mathbf{y}^{[m]} \in \mathbb{R}^{n}$, there exist matrices $\mathbf{J}_{1},\mathbf{J}_{2},\cdots,\mathbf{J}_{m} \in \mathbb{R}^{n\times n}$ such that
\begin{equation}
\begin{aligned}
    \label{eq:lemma-mvt}
    \mathbf{o}(\mathbf{x}^{[1]},\mathbf{x}^{[2]},\cdots,\mathbf{x}^{[m]}) - \mathbf{o}(\mathbf{y}^{[1]},\mathbf{y}^{[2]},\cdots,\mathbf{y}^{[m]}) \\
= \sum_{j=1}^{m} \mathbf{J}_{j}(\mathbf{x}^{[j]}-\mathbf{y}^{[j]}),
\end{aligned}
\end{equation}
and
\begin{equation}
    \label{eq:lemma-j-bound}
    \|\mathbf{J}_{1}\| \leq \sqrt{n}C,\quad \|\mathbf{J}_{2}\| \leq \sqrt{n}C, \quad \cdots, \quad\|\mathbf{J}_{m}\| \leq \sqrt{n}C.
\end{equation}
\end{lem}

To prove \textbf{Theorem \ref{thm1}}, we denote \[\hat{\mathbf{d}}^{[t]}=\mathbf{d}^{[t]}(\mathbf{x}^*,\nabla f(\mathbf{x}^*),\mathbf{x}^*,-\nabla f(\mathbf{x}^*),\mathbf{x}^*,\nabla f(\mathbf{x}^*)).\]
Then \eqref{eq:l2o-nesterov-x} can be written as
\[
\begin{aligned}
 \mathbf{x}^{[t+1]} = & \mathbf{x}^{[t]} - \mathbf{d}^{[t]}(\mathbf{x}^{[t]},\nabla f(\mathbf{x}^{[t]}),\mathbf{x}^{[t+1]},\mathbf{g}^{[t+1]},\mathbf{y}^{[t]},\nabla f(\mathbf{y}^{[t]}))\\
& + \mathbf{d}^{[t]}(\mathbf{x}^*,\nabla f(\mathbf{x}^*),\mathbf{x}^*,-\nabla f(\mathbf{x}^*),\mathbf{x}^*,\nabla f(\mathbf{x}^*)) - \hat{\mathbf{d}}^{[t]}.
\end{aligned}
\]
Applying Lemma~\ref{lemma:mvt}, we have
\[
    \begin{aligned}
 \mathbf{x}^{[t+1]} = \mathbf{x}^{[t]} & - \mathbf{J}_{1}^{[t]}(\mathbf{x}^{[t]}-\mathbf{x}^*) - \mathbf{J}_{2}^{[t]}(\mathbf{x}^{[t+1]}-\mathbf{x}^*) \\
& - \mathbf{J}_{3}^{[t]}(\mathbf{y}^{[t]}-\mathbf{x}^*)- \hat{\mathbf{d}}^{[t]}\\
& - \mathbf{J}_{4}^{[t]}(\nabla f(\mathbf{x}^{[t]}) - \nabla f(\mathbf{x}^*) ) \\
& - \mathbf{J}_{5}^{[t]}(\mathbf{g}^{[t+1]} + \nabla f(\mathbf{x}^*)) \\
& - \mathbf{J}_{6}^{[t]}(\nabla f(\mathbf{y}^{[t]}) - \nabla f(\mathbf{x}^*)),
\end{aligned}
\]
where matrices $\mathbf{J}_{j}^{[t]}(1 \leq j \leq 6)$ satisfy
\[\|\mathbf{J}_{j}^{[t]}\| \leq \sqrt{2N_a N_t}C, \quad \forall j=1,2,3,4,5,6.\]
Then, we perform some calculations and obtain
\[\begin{aligned}
 \mathbf{x}^{[t+1]} = \mathbf{x}^{[t]} & - \mathbf{J}_{1}^{[t]}(\mathbf{x}^{[t]}-\mathbf{x}^*) - \mathbf{J}_{2}^{[t]}(\mathbf{x}^{[t+1]}-\mathbf{x}^*) \\
& - \mathbf{J}_{3}^{[t]}(\mathbf{y}^{[t]}-\mathbf{x}^*)- \hat{\mathbf{d}}_{k}\\
& - (\mathbf{J}_{4}^{[t]} - \mathbf{J}_{5}^{[t]} + \mathbf{J}_{6}^{[t]} )(\nabla f(\mathbf{x}^{[t]}) - \nabla f(\mathbf{x}^*) ) \\
& - (\mathbf{J}_{5}^{[t]} - \mathbf{J}_{6}^{[t]})(\nabla f(\mathbf{x}^{[t]}) - \nabla f(\mathbf{x}^*) )\\
& - \mathbf{J}_{5}^{[t]}(\mathbf{g}^{[t+1]} + \nabla f(\mathbf{x}^*)) - \mathbf{J}_{6}^{[t]}(\nabla f(\mathbf{y}^{[t]}) - \nabla f(\mathbf{x}^*))\\
 = \mathbf{x}^{[t]} & - \mathbf{J}_{1}^{[t]}(\mathbf{x}^{[t]}-\mathbf{x}^*) - \mathbf{J}_{2}^{[t]}(\mathbf{x}^{[t+1]}-\mathbf{x}^*) \\
& - \mathbf{J}_{3}^{[t]}(\mathbf{y}^{[t]}-\mathbf{x}^*)- \hat{\mathbf{d}}^{[t]}\\
& - (\mathbf{J}_{4}^{[t]} - \mathbf{J}_{5}^{[t]} + \mathbf{J}_{6}^{[t]} )(\nabla f(\mathbf{x}^{[t]}) - \nabla f(\mathbf{x}^*) ) \\
& - (\mathbf{J}_{5}^{[t]} - \mathbf{J}_{6}^{[t]}) \nabla f(\mathbf{x}^{[t]}) - \mathbf{J}_{5}^{[t]}~ \mathbf{g}^{[t+1]} - \mathbf{J}_{6}^{[t]} \nabla f(\mathbf{y}^{[t]}).
\end{aligned}\]
Given any $\mathbf{B}^{[t]} \in \mathbb{R}^{2N_aN_t \times 2N_aN_t}$, as defined in \eqref{MS4L2O}, let
\[
\begin{aligned}
 \mathbf{P}_{1}^{[t]} &= \mathbf{J}_{5}^{[t]},\\
 \mathbf{P}_{2}^{[t]} &= \mathbf{J}_{6}^{[t]},\\
 \mathbf{b}_{1}^{[t]} &= \mathbf{J}_{1}^{[t]}(\mathbf{x}^{[t]}-\mathbf{x}^*) + \mathbf{J}_{2}^{[t]}(\mathbf{x}^{[t+1]}-\mathbf{x}^*) \\
 &\quad + \mathbf{J}_{3}^{[t]}(\mathbf{y}^{[t]}-\mathbf{x}^*) + \hat{\mathbf{d}}^{[t]} \\
 &\quad + (\mathbf{J}_{4}^{[t]} - \mathbf{J}_{5}^{[t]} + \mathbf{J}_{6}^{[t]} )(\nabla f(\mathbf{x}^{[t]}) - \nabla f(\mathbf{x}^*) ) \\
& \quad +\mathbf{B}^{[t]}(\mathbf{y}^{[t]} - \mathbf{x}^{[t]}). 
\end{aligned}
\]
Then we have
\[
\begin{aligned}
\mathbf{x}^{[t+1]} = &~ \mathbf{x}^{[t]} - (\mathbf{P}_{1}^{[t]}-\mathbf{P}_{2}^{[t]})\nabla f(\mathbf{x}^{[t]}) - \mathbf{P}_{2}^{[t]}\nabla f(\mathbf{y}^{[t]}) \\
& - \mathbf{P}_{1}^{[t]} ~\mathbf{g}^{[t+1]} + \mathbf{B}^{[t]}(\mathbf{y}^{[t]} - \mathbf{x}^{[t]}) - \mathbf{b}_{1}^{[t]},
\end{aligned}
\]
which exactly echos with \eqref{MS4L2O}. The upper bounds of $\mathbf{J}_{j}^{[t]}(1 \leq j \leq 6)$ imply that $\mathbf{P}_{1}^{[t]},\mathbf{P}_{2}^{[t]}$ are bounded, i.e.,
\[\|\mathbf{P}_{1}^{[t]}\| \leq \sqrt{2N_aN_t}C,\quad \|\mathbf{P}_{2}^{[t]}\| \leq \sqrt{2N_aN_t}C,\]
and $\mathbf{b}_{1}^{[t]}$ is controlled by
\begin{equation}\label{eq:b1k}
\begin{aligned}
 \|\mathbf{b}_{1}^{[t]}\| \leq &~ \sqrt{2N_aN_t}C \Big(\|\mathbf{x}^{[t]}-\mathbf{x}^*\| + \|\mathbf{x}^{[t+1]}-\mathbf{x}^*\| \\
& + \|\mathbf{y}^{[t]}-\mathbf{x}^*\| \Big) + \|\hat{\mathbf{d}}^{[t]}\| + \|\mathbf{B}^{[t]}\|\|\mathbf{y}^{[t]}-\mathbf{x}^{[t]}\| \\
& + 3\sqrt{2N_aN_t}C\|\nabla f(\mathbf{x}^{[t]}) - \nabla f(\mathbf{x}^*)\|.
\end{aligned}
\end{equation}
Since we set $\|\mathbf{b}_{1}^{[t]}\| \to 0$ as $t \to \infty$, according to \eqref{eq:b1k}, we have
\begin{equation}
\begin{aligned}
\|\mathbf{x}^{[t]}-\mathbf{x}^*\| \to 0,~~ \|\mathbf{x}^{[t+1]}-\mathbf{x}^*\|\to 0,\\
\|\mathbf{y}^{[t]}-\mathbf{x}^*\|\to 0,~~\|\mathbf{y}^{[t]}-\mathbf{x}^{[t]}\| \to 0,
\end{aligned}
\end{equation}
The proof for \textbf{Theorem \ref{thm1}} is thus completed.
\end{IEEEproof}

\bibliographystyle{IEEEtran}
\bibliography{IEEEabrv,learning}

\end{document}